\def\dir{./}
\documentclass[useAMS,usenatbib]{\dir mn2e}
\usepackage[fleqn]{amsmath}
\usepackage{amssymb}
\usepackage[super]{nth}
\usepackage{natbib}
\usepackage{graphicx}
\usepackage{float}
\usepackage{mathrsfs}
\usepackage{mathtools}
\usepackage{multirow}
\usepackage[usenames,dvipsnames]{color}
\usepackage{\dir aas_macros}
\usepackage{comment}
\usepackage{subfigure}
\usepackage{hyperref}
\usepackage{mathrsfs}
\usepackage{epstopdf, dcolumn}
\usepackage{wasysym}
\usepackage{dashrule}
\usepackage{xcolor}
\usepackage{textcomp}
\usepackage{threeparttable}

\definecolor{color_meraxes_line_1}{RGB}{27,158,119}
\definecolor{color_meraxes_line_2}{RGB}{217,95,2}
\definecolor{color_smaug_line_1}{RGB}{117,112,179}
\definecolor{color_smaug_line_2}{RGB}{231,41,138}

\definecolor{color_meraxes_fiducial}{RGB}{230,171,2}
\definecolor{color_meraxes_S}{RGB}{102,166,30}
\definecolor{color_meraxes_SM}{RGB}{231,41,138}
\definecolor{color_meraxes_SMB}{RGB}{217,95,2}
\definecolor{color_meraxes}{RGB}{27,158,119}

\newcommand{\appropto}{\mathrel{\vcentre{
 \offinterlineskip\halign{\hfil$##$\cr
 \propto\cr\noalign{\kern2pt}\sim\cr\noalign{\kern-2pt}}}}}
\newcommand{\Rom}[1]{\uppercase\expandafter{\romannumeral #1}}
\newcommand{\rom}[1]{\lowercase\expandafter{\romannumeral #1}}
\newcommand{\meraxes}{\textsc{Meraxes}}
\newcommand{\smaug}{{\textit{Smaug}}}
\newcommand{\nbody}{\textit{N}-body}
\newcommand{\htwo}{\mathrm{H}_2}
\newcommand{\hone}{\mathrm{H}\textsc{i}}

\defcitealias{Mutch2016a}{Paper-\Rom{3}}
\defcitealias{qin2018}{Paper-\Rom{14}}

\title[DRAGONS \Rom{15}: dwarf galaxies]{\LARGE Dark-ages~Reionization~and~Galaxy~Formation~Simulation~-~\Rom{15}.\\ Stellar evolution and feedback in dwarf galaxies at high redshift}
\author[Qin et al.]{Yuxiang Qin$^{1,2}$\thanks{E-mail: Yuxiang.L.Qin@Gmail.com}, Alan R. Duffy$^{3,2}$, Simon J. Mutch$^{1,2}$, Gregory B. Poole$^{1,3}$,
	\newauthor Andrei Mesinger$^4$ and J. Stuart B. Wyithe$^{1,2}$\\
	$^{1}$School of Physics, University of Melbourne, Parkville, VIC 3010, Australia\\
	$^{2}$ARC Centre of Excellence for All Sky Astrophysics in 3 Dimensions (ASTRO 3D)\\
	$^{3}$Centre for Astrophysics and Supercomputing, Swinburne University of Technology, PO Box 218, Hawthorn VIC 3122, Australia\\
	$^{4}$Scuola Normale Superiore, Piazza dei Cavalieri 7, I-56126 Pisa, Italy}

\begin{document}

%\date{Accepted 2014 December 2. Received 2014 December 1}
\date{\today\ draft - \nth{7}}
\pagerange{\pageref{firstpage}--\pageref{lastpage}} \pubyear{2018}
\maketitle
\label{firstpage}

\begin{abstract}
We directly compare predictions of dwarf galaxy properties in a semi-analytic model (SAM) with those extracted from a high-resolution hydrodynamic simulation. We focus on galaxies with halo masses of $10^9<M_\mathrm{vir}/\mathrm{M}_\odot\lesssim10^{11}$ at high redshift ($z\ge5$). We find that, with the modifications previously proposed in \citet{qin2018}, including to suppress the halo mass and baryon fraction, as well as to modulate gas cooling and star formation efficiencies, the SAM can reproduce the cosmic evolution of galaxy properties predicted by the hydrodynamic simulation. These include the galaxy stellar mass function, total baryonic mass, star-forming gas mass and star formation rate at $z\sim5-11$. However, this agreement is only possible by reducing the star formation threshold relative to that suggested by local observations. Otherwise, too much star-forming gas is trapped in quenched dwarf galaxies. We further find that dwarf galaxies rapidly build up their star-forming reservoirs in the early universe ($z>10$), with the relevant time-scale becoming significantly longer towards lower redshifts. This indicates efficient accretion in cold mode in these low-mass objects at high redshift. Note that the improved SAM, which has been calibrated against hydrodynamic simulations, can provide more accurate predictions of high-redshift dwarf galaxy properties that are essential for reionization study.
\end{abstract}


\section{Introduction}
Reionization refers to an important process after the Big Bang, during which the intergalactic medium (IGM) was transiting from neutral hydrogen to its ionized state \citep{Wyithe:2004kb}. According to the observed galaxy sample at high redshift \citep{Bouwens2014,Bouwens2015,Stefanon2016arXiv161109354S,Oesch2016ApJ...819..129O}, this process can only happen with ionizing photons from much fainter galaxies taken into account \citep{Robertson2013ApJ...768...71R,duffy2014low,Bouwens:2015hk,Liu2016}. Although there are still some debates on other possible sources that can dominate the high-redshift photon budget such as active galactic nuclei (AGN; \citealt{Giallongo2015A&A...578A..83G,Madau2015ApJ...813L...8M,Qin2017c,Hassan2018MNRAS.473..227H}), dwarf galaxies that are beyond our observational capabilities are generally thought to have driven the Epoch of Reionization (EoR). In this context, understanding the formation of these unobserved objects is crucial to studying the EoR and can only be probed at this stage with theoretical simulations. 

Hydrodynamic simulations evolve dark matter and baryonic particles simultaneously and provide direct insights into the relevant astrophysical process \citep{Vogelsberger2014,Schaye2014,Hopkins2014MNRAS.445..581H,feng2015bluetides}. However, resolving dwarf galaxies within a cosmological volume for reionization studies usually involves more than a few billions of particles, which remains computationally challenging at this stage. A more efficient method is to apply semi-analytic models (SAMs; \citealt{croton2006many,Somerville2008,guo2011dwarf,Henriques2015}) to \textit{N}-body simulations \citep{springel2005simulating,Boylan_Kolchin2009,Klypin2010,Garrison2017} that only consider collisionless particles. Using the halo properties inherited from the parent simulation, SAMs approximate baryonic physics such as gas accretion, star formation and feedback using simplified scaling relations. These relations are motivated directly from physical processes, or empirically from observational results and more complicated numerical techniques such as hydrodynamic simulations and radiative transfer calculations. The semi-analytic prescriptions that indirectly model galaxy formation introduce free parameters to describe efficiencies which are inevitably accompanied by parameter degeneracies \citep{Mutch2013,Henriques2013}. This can make their predictions sometimes controversial, and potentially disconnected from the \textit{true} behaviour in the universe.

An alternative to validate SAMs in the absence of observations at high redshift is to compare their results against hydrodynamic calculations that start from identical cosmological initial conditions. The goal of this work is to capture emergent behaviours from the hydrodynamic simulations (e.g. large-scale mass removal by winds from supernova events in individual star-forming sites) and to improve the parametrised modelling in SAMs as so to replicate these processes. Under the assumption that hydrodynamic simulations, which model the details of galaxy formation in a more physically realistic manner, are a more natural description of the astrophysical phenomenon and hence more representative of real galaxies, we can explore the semi-analytic prescriptions for quantities that are, in practice, unobservables and potentially reveal improper assumptions or missing physics in SAMs. 

\citet{Guo2016} compared the \textsc{l-galaxies} \citep{Cole2000,Bower2006} and \textsc{galform} \citep{Springel2005,Henriques2015} SAMs with the \textit{EAGLE} hydrodynamic simulations \citep{Schaye2014}, and concluded that the models can reproduce the stellar mass function predicted by \textit{EAGLE}. However, discrepancies were also found in the efficiencies of stellar and AGN feedback as well as the prediction of stellar mass-metallicity and size relations. \citet{Mitchell:2017je} also used \textit{EAGLE} to assess \textsc{galform} and found the angular momentum as well as the baryon cycling might not be properly traced in the SAM, leading to inaccurate predictions of galaxy sizes. \citet{Stevens:2017fi}, on the other hand, investigated cooling of Milk Way-like galaxies in \textit{EAGLE} and addressed the necessity of updating the cooling prescription employed in most SAMs (see recent updates of the cooling model in \citealt{Hou2018mnras.475..543h,Hou2018arxiv180301923h}). We, in the previous paper \citep{qin2018}, also found that the cooling prescription needs revision for more accurate modelling of \textit{low-mass} galaxies at \textit{high redshift} and proposed an alternative modification to the current prescription, avoiding the introduction of a new model. \citet{Cote:2017uh} recently extended the comparison from cosmological simulations of smoothed particles to zoom-in simulations \citep{Bryan2014ApJS..211...19B} of a system with a total mass of ${\sim}10^9\mathrm{M}_\odot$ (one main halo and two satellites; \citealt{Wise2012MNRAS.427..311W,Wise2012ApJ...745...50W,Wise2014MNRAS.442.2560W}), and investigated the difference in dwarf galaxy formation between a SAM and hydrodynamic simulation. They found their SAM, which employs a different prescription of gas accretion, was successful in reproducing the hydrodynamic calculation of star formation history but with a prediction of a much narrower distribution of metallicity compared to the hydrodynamic result.

This is the second paper following the work of \citet{qin2018}, where we investigated the performance of SAMs when applied to high-redshift dwarf galaxies. We used the {\meraxes} SAM \citep{Mutch2016a} as an example and focused on gas accretion, cooling and star formation with reionization and supernova feedback isolated. We compared the stellar and gas masses with a high-resolution hydrodynamic simulation from the {\smaug} suite \citep{duffy2010impact,duffy2014low}, and found that, in the SAM,
\begin{enumerate}
	\item due to the lack of hydrostatic pressure in parent \textit{N}-body simulations, inheriting halo properties directly from the dark matter halo merger trees overestimates the total mass of haloes hosting dwarf galaxies; 
	\item the assumption that, in the absence of feedback, haloes consists of a baryonic reservoir with a mass of $\Omega_\mathrm{b}/\Omega_\mathrm{m}$ of their total mass is not accurate for dwarf galaxy formation modelling and can lead to a significant overestimation of the total baryonic mass;
	\item star formation modelled by consuming the total gas disc in a few dynamical times of that disc cannot capture the evolutionary path of star formation implemented in hydrodynamic simulations; and
	\item gas accreted by dwarf galaxies is cold, the median temperature of which is significantly lower than the halo virial temperature, and the current cooling prescription is not representative of this process.
\end{enumerate}
Accordingly, we proposed modifications to SAMs, seeking for consistency with hydrodynamic simulations in calculations of the evolution of stellar and gas components of dwarf galaxies. In this work, we include these modifications as well as feedback from reionization and supernovae, and investigate whether the updated SAM can broadly agree with the hydrodynamic calculation of dwarf galaxies in the presence of feedback. 

We start with a brief review of the {\meraxes} SAM as well as the modifications proposed in \citet[hereafter \citetalias{qin2018}]{qin2018} and the {\smaug} hydrodynamic simulation suite in Section \ref{sec:models}. We then present and discuss our comparison results in Section \ref{sec:results}. Conclusions are given in Section \ref{sec:conclusion}. In this work, we adopt the Chabrier initial mass function (IMF; \citealt{chabrier2003galactic}) in the mass range of $0.1-120\mathrm{M}_\odot$ and cosmological parameters from \textit{WMAP7} ($\Omega_{\mathrm{m}}, \Omega_{\mathrm{b}}, \Omega_{\mathrm{\Lambda}}, h, \sigma_8, n_s $ = 0.275, 0.0458, 0.725, 0.702, 0.816, 0.968; \citealt{Komatsu2011}) in all simulations.

\section{The Dragons Project}\label{sec:models}
Taking advantage of \textit{N}-body/hydrodynamic simulations \citep{Poole2016,duffy2014low,Qin2017a} and SAMs \citep{Mutch2016a,Qin2017c}, the Dark-ages Reionization And Galaxy Formation Observable from Numerical Simulations (DRAGONS\footnote{\href{http://dragons.ph.unimelb.edu.au/}{http://dragons.ph.unimelb.edu.au}}) programme studies reionization and high-redshift galaxy formation \citep{Geil2016,Geil2017,Liu2016,Liu2017,Mutch2016b,Park2017,Duffy2017,Qin2017b}. In the previous publication of this series \citepalias{qin2018}, we used the {\meraxes} SAM as an example and investigated the semi-analytic modelling prescriptions adopted in the literature. We focussed on comparisons of dwarf galaxy properties calculated by {\meraxes} with a simplified model that ignores feedback from the {\smaug} hydrodynamic simulation suite. Based on the comparison, we proposed modifications to the halo properties as well as the cooling and star formation prescriptions.

In this work, we include reionization and supernova feedback, and extend the comparison of high-redshift ($z\ge5$) dwarf galaxy modelling with a molecular-hydrogen-based star formation law in the SAM. We briefly introduce the {\meraxes} SAM and {\smaug} hydrodynamic simulation in this section with emphasis on reionization and supernova feedback, and refer the interested reader to \citet[hereafter \citetalias{Mutch2016a}]{Mutch2016a}, \citet{Qin2017c} and Mutch et al. (in prep.) for more details of {\meraxes}, and \citet{Schaye2010} and \citet{duffy2014low} for the hydrodynamic simulations.

\subsection{\textsc{Meraxes}}\label{sec:meraxes}
{\meraxes} evolves galaxies with scaling relations capturing baryonic processes. These include gas infall, cooling, star formation, supernova feedback, metal enrichment, stellar mass recycling, reionization, supermassive black hole growth, AGN feedback\footnote{AGN feedback is not included in this work.} and mergers\footnote{Starburst driven by mergers is not discussed in this work while the relevant parameters are set to follow \citetalias{Mutch2016a}.}. It also calculates the ionization state of the IGM using the 21cm{\sc fast} semi-numerical algorithm \citep{Mesinger2010}. Note that, in order to take hydrostatic pressure into account \citepalias{qin2018} in this work, halo masses inherited from the merger trees that are constructed from collisionless {\nbody} simulations as well as total baryonic masses are updated using the halo mass and baryon fraction modifiers provided in \citet[][see Appendix \ref{sec:halo mass} for the impact of incorporating the modifiers to semi-analytic results]{Qin2017a}. We next provide a review of the star formation, supernova feedback and reionization feedback prescriptions in this section.

\subsubsection{Star formation}\label{sec:SAM_SF}
Gas falls into a halo from the IGM and cools through thermal radiation. Within $t_\mathrm{transition}$ (see Appendix \ref{sec:dis_gas_transition} for a review of the cooling prescription), this process leads to the formation of a star-forming disc ($m_\mathrm{sf}$), which is assumed to follow an exponential surface density profile
\begin{equation}\label{eq:sigma_H}
\Sigma_\mathrm{sf}(r) = \dfrac{m_\mathrm{sf}}{2\mathrm{\pi} r_\mathrm{disc}^2}\exp\left(-\dfrac{r}{r_\mathrm{disc}}\right),
\end{equation}
where $r_\mathrm{disc}{=} R_\mathrm{vir}\dfrac{\lambda}{\sqrt{2}}$ ($\lambda$ is the halo spin parameter as defined in \citealt{bullock2001apj...555..240b}) represents the disc scalelength assuming the conservation of specific angular momentum between the star-forming disc and its host halo \citep{Mo1998}. 

\textbf{Total-gas-based star formation prescription:} In previous DRAGONS publications, we follow \citet{croton2006many}, form stars by consuming the total star-forming gas, and calculate the star formation rate (SFR) by
\begin{equation}\label{eq:sfr} 
\dot{m}_* = \dfrac{\max\left(0, m_\mathrm{sf}-m_\mathrm{sf,c}\right)}{t_\mathrm{sf}},
\end{equation}
where $t_\mathrm{sf}\equiv \alpha^{-1}_\mathrm{sf}3r_{\mathrm{disc}}/V_\mathrm{vir}$ represents the depletion time-scale of total star-forming gas on the disc. Note that $3r_{\mathrm{disc}}$ corresponds to the outer disc radius where star formation can happen according to observations of the Milky Way \citep{2000glg..book.....V}. $\alpha_{\mathrm{sf}}$, $V_{\mathrm{vir}}$ and $m_\mathrm{sf,c}$ are the star formation efficiency, virial velocity and the minimum mass of hydrogen gas that a galaxy can form stars, respectively. The form of $t_\mathrm{sf}$ indicates that the depletion time-scale of total star-forming gas approximately follows the dynamical time-scale of the host halo. However, this was found to be inconsistent with the implementation in hydrodynamic simulations \citepalias{qin2018}. In this work, we instead adopt the following equation and use free parameters, $\alpha_\mathrm{sf}$ and $\beta_\mathrm{sf}$, to directly adjust the evolutionary path of $t_\mathrm{sf}$
\begin{equation}\label{eq:t_sfr}
t_\mathrm{sf} = \alpha_\mathrm{sf}\left(\dfrac{1+z}{6}\right)^{-\beta_\mathrm{sf}}.
\end{equation}

\textbf{Molecular-hydrogen-based star formation prescription:} A second star formation prescription will be explored in this work, the detail of which will be presented in Mutch et al. (in prep.). Note that this prescription is based on the depletion of molecular hydrogen (see \citealt{Lagos2011} and references therein) and is considered as a more physically plausible model compared to the total-gas-based star formation prescription. First, we assume stars ($m_*$) follow the same distribution as the interstellar medium (ISM) and they are considered as a stellar disc with a surface density profile following equation (\ref{eq:sigma_H}; changing the subscript from $_\mathrm{sf}$ to $_*$). We then calculate the pressure of the ISM accounting for both stars and hydrogen through the \citet{Elmegreen1993} approximation
\begin{equation}
P_\mathrm{ISM}\left(r\right) \approx 0.5\mathrm{\pi} G \Sigma_\mathrm{sf}\sigma_\mathrm{sf}\left[\dfrac{\Sigma_\mathrm{sf}}{\sigma_\mathrm{sf}}+\dfrac{\Sigma_{*}}{\sigma_*}\right],
\end{equation}
where $G$ represents the gravitational constant and, following \citet{Lagos2011}, the vertical velocity dispersions of gas ($\sigma_\mathrm{sf}$) and stars ($\sigma_*$) are assumed to be $10\mathrm{km\ s^{-1}}$ and $0.02\sqrt{r_\mathrm{disc} \mathrm{\pi} G \Sigma_{*}}$, respectively. In order to split the disc into molecules and atoms, the observed relation between the ISM pressure and the surface density ratio of $\hone$ to $\htwo$ is implemented. \citet{Blitz2006} investigated the $\hone$, CO and stellar densities of 14 nearby galaxies and found 
\begin{equation}
\dfrac{\Sigma_{\hone}\left(r\right)}{\Sigma_{\htwo}\left(r\right)} =  \left(\frac{P_\mathrm{ISM}\left(r\right)/k_\mathrm{B}}{10^{4.54\pm0.07}\mathrm{cm}^{-3}\mathrm{K}}\right)^{0.92{\pm}0.07},
\end{equation}
where $k_\mathrm{B}$ represents the Boltzmann constant. With $\Sigma_{\htwo}\left(r\right)+\Sigma_{\hone}\left(r\right)=\Sigma_\mathrm{sf}\left(r\right)$, we estimate the $\htwo$ mass according to
\begin{equation}\label{eq:H2}
m_{\htwo} = \min\left[m_\mathrm{sf}, \int^{5r_\mathrm{disc}}_{0}\Sigma_{\htwo}\left(r\right)2\mathrm{\pi} r \mathrm{d}r\right]
\end{equation}
and calculate the SFR following equations (\ref{eq:sfr}) and (\ref{eq:t_sfr}; changing the subscript from $_\mathrm{sf}$ to $_\mathrm{\htwo}$). Note that we do not impose any mass threshold of $\htwo$ in this work and set $m_\mathrm{\htwo,c}=0$.

\subsubsection{Supernova feedback}\label{sec:SAM_SN}
Inferred from the stellar lifetime-mass relation \citep{Portinari1997} and the assumed IMF, we calculate the fraction of the newly formed stars ($\Delta m_\star=\dot{m}_\star\Delta t$ where $\Delta t$ is the time interval between two snapshots) which will have reached the supernova stage at the end of the current time step, $\eta_\mathrm{SNII}$. These stars recycle their masses to the ISM, and the metals and energy produced by the supernovae provide feedback to the environment. In particular, the metals enhance the cooling rate through the metallicity dependent cooling function while the supernova energy leads to transition of gas between different reservoirs. In practice, supernova energy converts star-forming gas\footnote{$\hone$ and $\htwo$ are not distinguished when applying supernova feedback. We keep distributing supernova energy and metals to the entire star-forming gas reservoir.} ($m_\mathrm{sf}$) to hot (i.e. non-star-forming gas; $m_\mathrm{hot}$) and, in the case of strong supernova feedback, further ejects hot gas ($m_\mathrm{ejected}$) from the galaxy.

In this work, we adopt the \citet{guo2011dwarf} prescriptions to calculate the energy coupled to the surrounding gas
\begin{equation}\label{eq:esn}
e_{\mathrm{total}} = \epsilon_\mathrm{energy}\eta_\mathrm{SNII}\Delta m_{\star}\times10^{51}\mathrm{erg},
\end{equation}
with
\begin{equation}\label{eq:epsilon_energy}
\epsilon_\mathrm{energy} = \mathrm{min}\left\{1, \alpha_{\mathrm{energy}}\left[\dfrac{1}{2}+\left(\dfrac{V_\mathrm{max}}{V_\mathrm{energy}}\right)^{-\beta_{\mathrm{energy}}}\right]\right\}
\end{equation}
representing the coupling efficiency between supernova energy and the ISM, where $V_\mathrm{max}$ is the maximum circular velocity of the host halo, $\alpha_{\mathrm{energy}}$, $\beta_{\mathrm{energy}}$ and $V_{\mathrm{energy}}$ are free parameters introduced to modulate the feedback efficiency.

We consider the following two supernova feedback regimes: 1) \textit{contemporaneous feedback} which is provided by massive stars formed within the current snapshot; and 2) \textit{delayed feedback} where long-lived stars formed at earlier times are taken into account. Therefore, the total energy released in the current snapshot, $j$, is 
\begin{equation}\label{eq:Esn}
E_{\mathrm{total}} {=} \mathrm{\sum}^{i=j}_{i=j-4} e_{\mathrm{total},i}^{j}\left(\Delta m_{{\star},i}^{j}, \epsilon_{\mathrm{energy}}^{j}, \eta_{\mathrm{SNII},i}^{j}\right),
\end{equation}
where $j-4$ represents \textit{delayed feedback} from the previous 4 snapshots, and $e_{\mathrm{total},i}^{j}\left(\Delta m_{{\star},i}^{j}, \epsilon_{\mathrm{energy}}^{j}, \eta_{\mathrm{SNII},i}^{j}\right)$ denotes the supernova energy ejected by stars that are formed at snapshot $i$ and become supernovae at snapshot $j$.

On the other hand, the expected mass of gas that is heated by supernovae depends on the mass loading factor, $\epsilon_{\mathrm{mass}}$, which is assumed to follow the same form as the coupling efficiency in equation (\ref{eq:epsilon_energy}; changing the subscript from $_\mathrm{energy}$ to $_\mathrm{mass}$). We first calculate the maximum reheated mass by
\begin{equation}\label{eq:m_reheat}
\Delta m_{\mathrm{sf}}^\mathrm{max} = \mathrm{\sum}^{i=j}_{i=j-4} \epsilon_{\mathrm{mass}}^{j} \eta_{\mathrm{SNII},i}^{j}\Delta m_{{\star},i}^{j},
\end{equation}
with an upper limit of $\epsilon_\mathrm{mass}$ set to be $\epsilon_\mathrm{mass}^\mathrm{max}=10$ following \citetalias{Mutch2016a}, which is a typical value for high-redshift dwarf starburst galaxies \citep{uhlig2012}. Note that, since the total supernova energy is finite (see equation \ref{eq:Esn}), a galaxy might not be able to heat all the mass estimated from the loading factor. Therefore, we calculate the mass of actually reheated gas according to
\begin{equation}\label{eq:Delta m_reheat}
\Delta m_{\mathrm{sf}} = -\mathrm{min}\left[m_\mathrm{sf}, \mathrm{min}\left(\dfrac{E_{\mathrm{total}}}{0.5V_\mathrm{vir}^{2}}, \Delta m_{\mathrm{sf}}^\mathrm{max}\right)\right].
\end{equation}
While in the case of an intense supernova event where $E_{\mathrm{total}}+0.5\Delta m_\mathrm{sf}V_\mathrm{vir}^{2}>0$, the energy released by supernovae further unbinds the hot gas which is removed from the galaxy and stored in a reservoir termed the ejected gas
\begin{equation}\label{eq:m_eject}
\Delta m_{\mathrm{hot}}= -\mathrm{min}\left[m_\mathrm{hot}, \mathrm{max}\left(0, \dfrac{E_{\mathrm{total}}}{0.5V_\mathrm{vir}^{2}} + \Delta m_\mathrm{sf}\right)\right].
\end{equation}
We assume the ejected gas does not contribute to star formation due to its low-density and high-temperature profile, and its cooling process might only be efficient when it has been reincorporated\footnote{Reincorporating gas that has been ejected by supernova feedback is not included in this work.} into the galaxy \citep{Henriques2013}.

\subsubsection{Reionization feedback}\label{sec:SAM_RE}
In order to model the feedback from reionization, we further\footnote{Two baryon fraction modifiers are considered in this work, corresponding to reionization feedback and hydrostatic pressure.} inhibit the local baryon fraction of haloes by a factor of 
\begin{equation}\label{eq:f_mod_meraxes}
 f_\mathrm{mod}\equiv2^{-{M}_\mathrm{crit}/M_\mathrm{vir}},
\end{equation}
where $M_\mathrm{vir}$ is the halo mass and ${M}_\mathrm{crit}$ represents a filtering mass below which haloes are not able to efficiently accrete baryons from the IGM. We calculate the critical mass for each halo following \citet{Sobacchi2013a}
\begin{equation}\label{eq:M_crit}
\dfrac{M_\mathrm{crit}\left(\textbf{r},z\right)}{2.8{\times}10^9\mathrm{M}_\odot} {=} J_{21}^{0.17}\left(\textbf{r},z\right)\left(\dfrac{1{+}z}{10}\right)^{{-}2.1}\left[1{-}\left(\dfrac{1+z}{1{+}z_\mathrm{ion}}\right)^{2.0}\right]^{2.5},
\end{equation}
where $J_{21}\left(\textbf{r},z\right)$ represents the local UV background intensity while $z_\mathrm{ion}$ is the redshift at which the surrounding IGM was first ionized and is determined using the 21cm{\sc fast} algorithm (\citealt{Mesinger2010}; see \citetalias{Mutch2016a} for more details).

\subsection{{\smaug}}\label{sec:smaug}
{\smaug}, a high-resolution hydrodynamic simulation suite, was performed using a modified version of the {\sc gadget}-2 \textit{N}-body/hydrodynamics code \citep{springel2005cosmological}, following the same parameter configuration of the OverWhelmingly Large Simulations project (OWLS; \citealt{Schaye2010}). The simulations presented in this work start from the same initial conditions generated with the {\sc grafic} package \citep{Bertschinger2001} at $z=199$ using the Zel'dovich approximation \citep{Zeldovich1970}. Each simulation evolves $\left(2\times\right)512^3$ particles including dark matter (and baryons) within a periodic cube of comoving side of $10h^{-1}\mathrm{Mpc}$. The Plummer-equivalent comoving softening length is $0.2h^{-1}\mathrm{kpc}$ and the particle resolution is 4.7 and $0.9\times10^5h^{-1}\mathrm{M}_\odot$ for dark matter and baryons or $5.7\times10^5h^{-1}\mathrm{M}_\odot$ if only dark matter particles are considered. We summarize the adopted subgrid physics prescriptions in this section.

\begin{enumerate}
	\item \textbf{Cooling} \citep{Wiersma2009a} consists of both primordial elements and metal emission lines from carbon, nitrogen, oxygen, neon, magnesium, silicon, sulphur, calcium and iron. The cooling function is pre-tabulated using the {\sc cloudy} package \citep{ferland1998cloudy}, and accounts for free--free scattering between gas particles as well as Compton scattering between gas particles and cosmic microwave background (CMB) photons.
	\item \textbf{Star formation} \citep{Schaye2008} occurs in the ISM that is assumed to be multiphase. The following equation describe its state
	\begin{equation}
	\dfrac{T_\mathrm{g,eos}}{8\times10^3\mathrm{K}}  =\left(\dfrac{n_\mathrm{H}}{10^{-1}\mathrm{cm}^{-3}}\right)^{\gamma_\mathrm{eff}-1},
	\end{equation}
	where $T_\mathrm{g,eos}$, $n_\mathrm{H}$ and $\gamma_\mathrm{eff}=4/3$ represent the gas temperature on the equation of state (EOS), total hydrogen number density and effective ratio of specific heats, respectively. 
	
	Gas particles are considered as on the EOS and identified as potentially star-forming regions when they become dense and cold. Star formation is then implemented by stochastically converting star-forming gas particles to star particles with a probability of $\min\left(1, m_\mathrm{g}\Delta t/t_\mathrm{g}\right)$, where $t_\mathrm{g}$ represents the gas depletion time-scale. In the case that the ISM has formed a self-gravitating disc, its scaleheight is of the order of the local Jeans length. Based on this, the depletion time-scale follows
	\begin{equation}
	t_\mathrm{g} \equiv \dfrac{\dot{m}_*}{m_\mathrm{g}} \approx 1.67\mathrm{Gyr}\left(\dfrac{P_\mathrm{g}/k_\mathrm{B}}{10^3\mathrm{cm}^{-3}\mathrm{K}}\right)^{-0.2}, %A^{-1}\left(1\mathrm{M}_\odot\mathrm{pc}^{-2}\right)^n\left(\dfrac{f_\mathrm{g}P_\mathrm{g}/k_\mathrm{B}}{10^3\mathrm{cm}^{-3}\mathrm{K}}\right)^{\left(n-1\right)/2}
	\end{equation}	
	the normalization and scaling of which were determined by observations -- the KS law \citep{kennicutt1998global}.
	
	\item \textbf{Supernova feedback} can be simulated kinetically \citep{DallaVecchia2008} with supernova winds pushing nearby gas particles at given probabilities and velocities representing efficiencies, or can be modelled thermally \citep{DallaVecchia2012} by stochastically distributing supernova energy to the surrounding gas particles and increasing the gas temperature by a given amount ($\Delta T = 10^{7.5}\mathrm{K}$). In order to compare galaxy properties between {\meraxes} and {\smaug}, we focus on the hydrodynamic simulation with thermal supernova feedback implemented, which behaves similarly to the semi-analytic supernova feedback prescription (see Section \ref{sec:SAM_SN}).
	
	Note that each star particle represents a single stellar population that is described by the Chabrier IMF\citep{chabrier2003galactic}, and its mass decreases due to stellar recycling, which accounts for winds from AGB and massive stars as well as Type Ia and II supernovae \citep{Wiersma2009b}. The total number of stars per stellar mass that will reach core-collapse supernovae at the end of their life cycle ($\eta_\mathrm{SNII}=1.19\times10^{-2}\mathrm{M}_\odot^{-1}$) is inferred from the IMF. The supernova energy produced by a star particle ($m_*$) is stochastically distributed onto its $N_\mathrm{ngb}=48$ nearby gas particles with a probability of $\min\left(1, \dfrac{10^{51} \mathrm{erg}\times f_\mathrm{th}\eta_\mathrm{SNII} m_*}{\Sigma_{i=1}^{N_\mathrm{ngb}}m_\mathrm{g,i}\Delta E_\mathrm{g}}\right)$, where $f_\mathrm{th}$ is the fraction of energy that contributes to feedback, which is set to be unity in {\smaug}; $m_\mathrm{g,i}$ represents the mass of gas particle $i$; and $\Delta E_\mathrm{g}$ corresponds to the energy increment when the particle temperature is raised by $\Delta T$.
	
	\item \textbf{Reionization feedback} is implemented as a UV/X-ray background \citep{Haardt2001} with all gas particles (assuming optically thin) being instantaneously heated to $10^4\mathrm{K}$ at a given redshift, $z_\mathrm{re}$. Although this prescription is numerically achievable and is considered appropriate in the $10h^{-1}$Mpc volume of {\smaug} simulations \citep{duffy2014low}, it is not an accurate calculation of reionization feedback. Therefore, the semi-analytic prescription of reionization feedback (see Section \ref{sec:SAM_RE}) will be compared to two hydrodynamic simulations with $z_\mathrm{re}=9$ and 6.5, which bracket the observed CMB and Ly-$\alpha$ forest boundaries of the EoR, and represent the strongest and weakest feedback scenarios, respectively.
\end{enumerate}

Simulations utilized in this study is summarized below:
\begin{enumerate}
	\item[(1)] \textit{DMONLY}, a collisionless \textit{N}-body simulation including only dark matter particles and neglecting baryonic physics. It is used to construct dark matter halo merger trees for running {\meraxes};
	\item[(2)] \textit{NOSN\_NOZCOOL}, \textit{NOSN\_NOZCOOL\_LateRe} and \textit{NOSN\_NOZCOOL\_NoRe}, three toy models with cooling in the absence of metal line emission and ignoring supernova feedback. While \textit{NOSN\_NOZCOOL\_NoRe} does not include reionization feedback, the UV/X-ray heating backgrounds of the first two simulations are switched on at $z_\mathrm{re}=9$ and 6.5, respectively. These three models will be used to compare with the SAM and to investigate the homogeneous reionization feedback prescription. 
	\item[(3)] \textit{WTHERM}, a complete\footnote{\textit{WTHERM} does not include AGN feedback either.} hydrodynamic simulation including radiative cooling from primordial elements and metals, as well as stellar evolution, thermal supernova feedback -- the energy produced by supernovae stochastically increases the temperature of the nearby ISM, and instantaneous photonionization heating from a reionization background at $z_\mathrm{re}=9$. We will use \textit{WTHERM} to investigate the supernova feedback recipe implemented in the SAM.
\end{enumerate} 

\section{Comparison between dwarf galaxies in SAM and hydrodynamic simulations}\label{sec:results}
In this work, with algorithms described in \citetalias{qin2018}, we 1) include the aforementioned modifications of the semi-analytic properties or prescriptions including the halo mass, baryon fraction, gas transition time-scale and star formation time-scale; 2) build halo merger trees using the \textit{DMONLY} simulation; 3) match each individual galaxies\footnote{Central galaxies with $M_\mathrm{vir}>10^{7.5}\mathrm{M}_\odot$ are considered as well-resolved in our simulation and included in this work.} between {\meraxes} and {\smaug} outputs; and 4) identify star-forming and hot gas in {\smaug}. We present the comparison result between dwarf galaxy properties predicted by the hydrodynamic simulation and SAM in this section.

\subsection{Reionization feedback}\label{sec:reionization feedback}
\begin{figure}
	\centering
	\includegraphics[width=0.92\columnwidth]{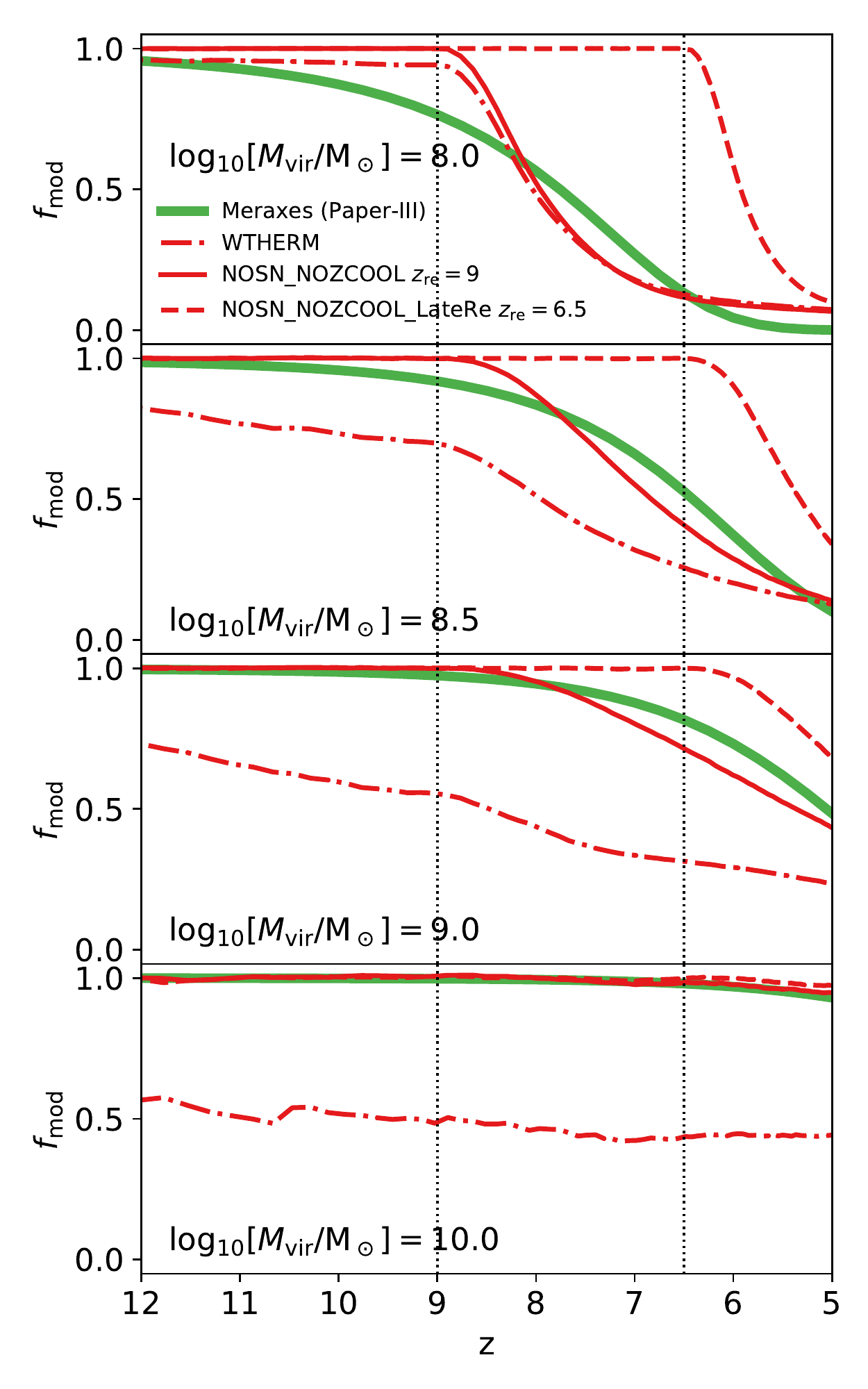}\vspace*{-3mm}
	\caption{\label{fig:reionization_modifiers}Baryon fraction modifiers due to reionization (not to be confused with the baryon fraction modifier due to hydrostatic pressure, see Appendix \ref{sec:halo mass}). The green solid thick line represents the homogeneous reionization background incorporated in {\meraxes}, which is proposed in \citetalias{Mutch2016a}. The red solid and dashed thin lines demonstrate the effect of the UV/X-ray background in the \textit{NOSN\_NOZCOOL} and \textit{NOSN\_NOZCOOL\_LateRe} {\smaug} simulations where gas particles are instantaneously heated to $10^4\mathrm{K}$ at $z_\mathrm{re}=9$ or 6.5 (marked by vertical dotted lines), respectively. The baryon fraction of halos in the \textit{WTHERM} simulation where $z_\mathrm{re}=9$ is shown by the red dash-dotted line, which indicates the suppression due to supernova feedback.}
\end{figure}

Reionization feedback in the SAM is incorporated by inhibiting the local baryon fraction of haloes using a filtering mass (see Section \ref{sec:SAM_RE} or \citetalias{Mutch2016a} for more details). In this work, we adopt the average filtering mass, $\bar{M}_\mathrm{crit}\left(z\right)$, proposed in \citetalias{Mutch2016a}, which ignores the spatial distribution of the IGM ionization state and depends only on redshift. In order to assess the validity of this feedback prescription\footnote{Note that \citetalias{Mutch2016a} compared this model with their fiducial calculation that adopts the 21cm{\sc fast} algorithm \citep{Mesinger2010} to self-consistently evaluate the local UV background and filtering mass ($\bar{M}_\mathrm{crit}\left(\textbf{r},z\right)$). They found the two models predict similar stellar mass build-up and global ionizing history.}, we compare with two {\smaug} hydrodynamic simulations in which gas particles are instantaneously heated to $10^4\mathrm{K}$ at $z_\mathrm{re}=9$ (\textit{NOSN\_NOZCOOL}) and 6.5 (\textit{NOSN\_NOZCOOL\_LateRe}), respectively (see Section \ref{sec:smaug}). Note that the suppression due to the ionizing background is quantified in the SAM using a baryon fraction modifier (see equation \ref{eq:f_mod_meraxes}), which, in hydrodynamic simulations, can be informed by comparing the baryonic components of galaxies matched between the \textit{NOSN\_NOZCOOL\_NoRe} and \textit{NOSN\_NOZCOOL(\_LateRe)} results \citep{Qin2017a}
\begin{equation}\label{eq:f_mod_smaug}
f_\mathrm{mod}=\dfrac{f_\mathrm{b}\left[NOSN\_NOZCOOL(\_LateRe)\right]}{f_\mathrm{b}\left(NOSN\_NOZCOOL\_NoRe\right)}.
\end{equation}
We show the the reionization modifiers adopted in {\meraxes} (equation \ref{eq:f_mod_meraxes}) and calculated from {\smaug} (equation \ref{eq:f_mod_smaug}) in Fig. \ref{fig:reionization_modifiers}. We see that, through photonionization heating, reionization plays a significant role in reducing the fraction of baryons, and that the baryon fraction modifier adopted in the SAM is in general agreement with the hydrodynamic result -- $f_\mathrm{mod}$ decreases in less massive haloes and towards lower redshifts.

\subsubsection{Reionization in the hydrodynamic simulations}
In a $10{\rm h^{-1}Mpc}$ volume, the UV/X-ray ionizing background adopted in the two hydrodynamic simulations represents the strongest ($z_\mathrm{re}=9$) and weakest ($z_\mathrm{re}=6.5$) feedback scenarios that are consistent with the CMB and Lyman $\alpha$ observations \citep{duffy2014low}. However, since reionization does not affect the gas component before $z_\mathrm{re}$ in these simulations, its feedback on the baryonic reservoirs cannot be captured by the simulations at higher redshifts. Therefore, the time when reionization feedback becomes important is relatively late compared to the SAM where the onset of reionization is more gradual and realistic \citep{Sobacchi2013a} on large scales. For instance, for halos with $M_\mathrm{vir}\sim10^8\mathrm{M}_\odot$, $z|_{f_\mathrm{mod}=0.9}$ is larger than 10 in {\meraxes} while it is around 8.5 and 6 in the two {\smaug} results. As a result, the baryon fraction is overestimated in the hydrodynamic simulations at earlier times, which can potentially lead to an overproduction of stellar mass in dwarf galaxies. 

\begin{table*}
	\caption{A list and description of the main SAM parameters used in this work. \textit{SAM\_KS\_(un)limited} and \textit{SAM\_H2} represent models using the total- and molecular-hydrogen-based star formation prescriptions, respectively, while the third row shows the parameters adopted for the fiducial model in \citetalias{Mutch2016a}\tnote{a} for comparison.}\label{tab:parameters}
	\begin{threeparttable}
		\begin{tabular}{p{25mm}|p{29mm}|p{27mm}|p{24mm}|p{25mm}|p{22mm}}
			\hline
			\hline\rule{0pt}{3ex}%\vspace*{1.5mm}
			Parameters&
			$t_{\mathrm{transition}}$\tnote{b}&
			$m_\mathrm{sf(\htwo),c}$&
			$t_{\mathrm{sf}(\htwo)}$&
			$\epsilon_{\mathrm{energy}}$\tnote{c}&
			$\epsilon_{\mathrm{mass}}$\tnote{c}\\\hline\rule{0pt}{3ex}
			
			Description&
			Time-scale of gas transition from hot to star-forming in Gyr&
			Minimum mass of gas\tnote{d,e} \ for star formation in $10^8\mathrm{M}_\odot$&
			Depletion time-scale of hydrogen\tnote{d} \ gas in Gyr&
			Coupling efficiency of supernovae energy and the ISM\tnote{e}&
			Mass loading factor for supernovae heating\\\hline\rule{0pt}{3ex}
			
			\textit{SAM\_PaperIII}&
			$0.18\left(\dfrac{1+z}{6}\right)^{-1.5}$&
			$1.4V_\mathrm{vir,70}\left(\dfrac{r_\mathrm{dics}}{1h^{-1}\mathrm{kpc}}\right)$&
			$0.41\left(\dfrac{1+z}{6}\right)^{-1.5}$&
			$0.5\left(\dfrac{1}{2}+V_\mathrm{max,70}^{-2}\right)$&
			$6$\\\hline\rule{0pt}{3ex}
			
			\textit{SAM\_KS\_limited}&
			$0.72\left(\dfrac{1+z}{6}\right)^{-3.5}$&
			$1.4V_\mathrm{vir,70}\left(\dfrac{r_\mathrm{dics}}{1h^{-1}\mathrm{kpc}}\right)$&
			$0.81\left(\dfrac{1+z}{6}\right)^{-0.2}$&
			$0.04\left(\dfrac{1}{2}+V_\mathrm{max,70}^{-2}\right)$&
			$1.5$\\\hline\rule{0pt}{3ex}
			
			\textit{SAM\_KS\_unlimited}&
			$18\left(\dfrac{1+z}{6}\right)^{-7.5}$&
			$0$&
			$4.1\left(\dfrac{1+z}{6}\right)^{-1.0}$&
			$0.02\left(\dfrac{1}{2}+V_\mathrm{max,70}^{-2}\right)$&
			$1.5$\\\hline\rule{0pt}{3ex}
			
			\textit{SAM\_H2}&
			$6.0\left(\dfrac{1+z}{6}\right)^{-6.5}$&
			$0$&
			$0.90\left(\dfrac{1+z}{6}\right)^{-0.8}$&
			$0.03\left(\dfrac{1}{2}+V_\mathrm{max,70}^{-2}\right)$&
			$1.5$\\\hline
			
			\hline
		\end{tabular}
		\begin{tablenotes}
			\item[a] The particle masses of the parent \textit{N}-body simulations are $8.1\times10^{5}\mathrm{M}_\odot$ in this work and $3.9\times10^{6}\mathrm{M}_\odot$ in \citetalias{Mutch2016a}, where the Chabrier and Salpeter IMFs are adopted, respectively.		
			\item[b] $t_\mathrm{transition}\ge0.2t_\mathrm{dyn}\equiv36\left[\left(1+z\right)/6\right]^{-1.5}\mathrm{Myr}$.
			\item[c] $\epsilon_\mathrm{energy}\le1$ and $\epsilon_\mathrm{mass}\le\epsilon_\mathrm{mass}^{max}\equiv10$.
			\item[d] Total star-forming gas and molecular hydrogen for \textit{SAM\_KS\_(un)limited} and \textit{SAM\_H2}, respectively.
			\item[e] $V_\mathrm{vir(max),70}\equiv V_\mathrm{vir(max)}/70\mathrm{km\ s^{-1}}$ where $V_\mathrm{vir}$ and $V_\mathrm{max}$ are the virial velocity and maximum circular velocity of host halos.		
		\end{tablenotes}
	\end{threeparttable}
\end{table*}

More accurate modelling of reionization requires dedicated calculations of radiative transfer, which is numerically consuming and hence a challenging task (see a review and comparison of cosmological radiative transfer codes by \citealt{Iliev2006} and a recent comparison by \citealt{hutter2018} between semi-numerical reionization modelling and radiative transfer). In the following sections, we will focus on galaxies in the range of $M_\mathrm{vir}>10^{9}\mathrm{M}_\odot$, where reionization plays a similar and insignificant role in suppressing baryon fractions in {\meraxes} and {\smaug}.

\subsection{Stellar evolution and feedback}
We next investigate the stellar evolution and feedback in {\meraxes} and {\smaug}, starting with a discussion of involved free parameters in the SAM.

\subsubsection{Degeneracy of the parameter space in SAMs}\label{sec:paperiii}

Cosmological SAMs are usually calibrated against the observed galaxy stellar mass functions (or equivalently the SFR functions or galaxy luminosity functions) where a sufficient sample is available. By doing this, the stellar component is assured to be well modelled in a statistical context, and with more upcoming observations, the parameter space becomes better constrained and missing physics in the SAM might be revealed (e.g. AGN feedback \citealt{croton2006many}). However, one of the issues about this calibration strategy is that it cannot guarantee the modelled galaxies are also representative of real galaxies in terms of their unobservable properties, considering most baryonic processes are modelled indirectly in SAMs with the relevant parameters poorly understood. Taking the gas component as an example, although a handful of radio telescopes are capable of observing the gas component of distant galaxies (e.g. ALMA\footnote{\url{http://www.almaobservatory.org}}; \citealt{Aravena:2016hb,Bradac:2017bf,Carniani:2017jp,Falgarone:2017if}), unfortunately, such tasks are still extremely challenging. The current sample of observed gas components of distant galaxies remains small, limiting our understanding of how galaxies accrete baryons and convert their hydrogen into stars in the early universe.

\begin{figure*}
	\begin{minipage}{1\textwidth}
		\centering
		\hspace*{-4mm}
		\includegraphics[height=13.2cm]{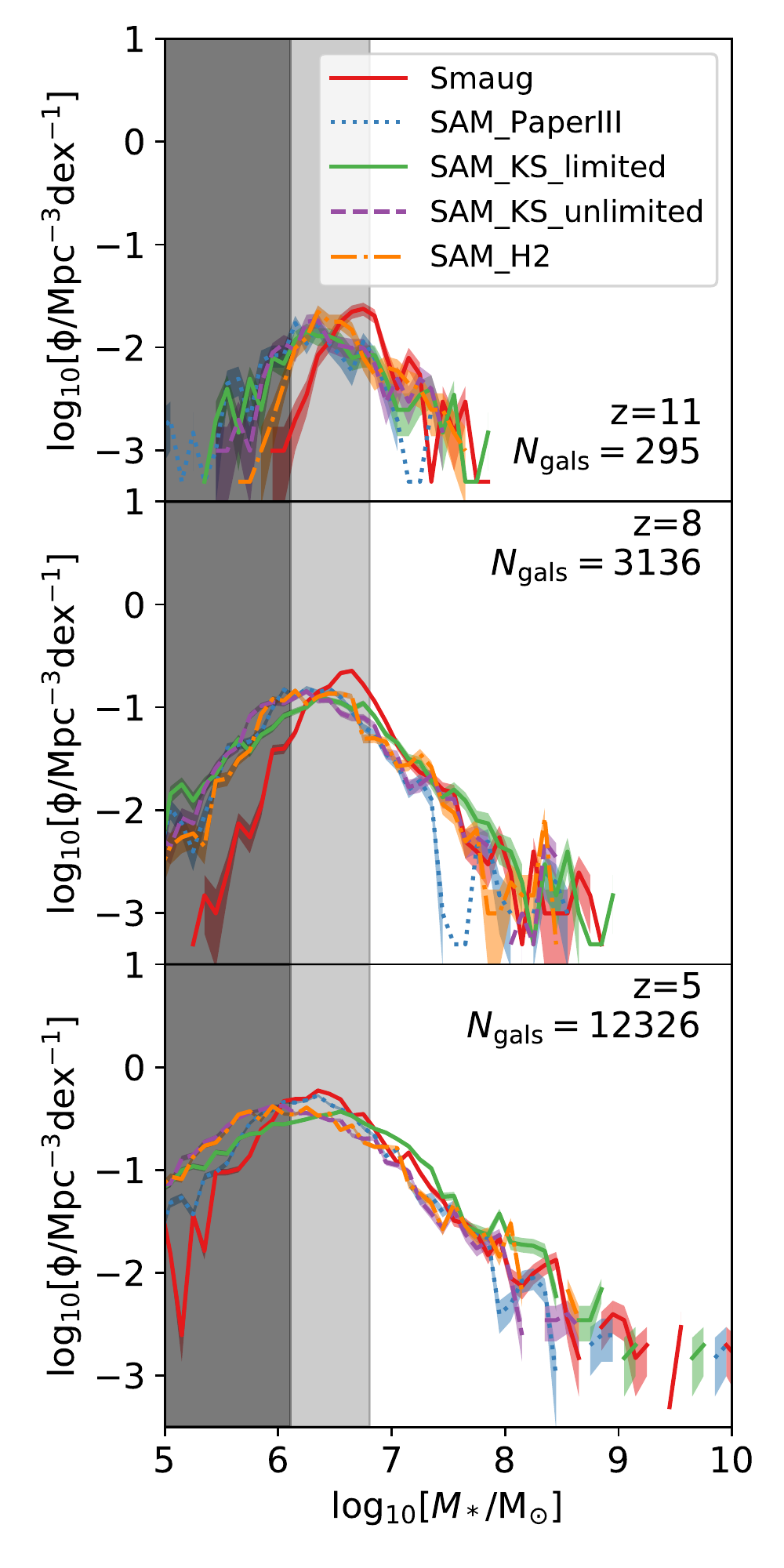}\hspace*{-4mm}
		\includegraphics[height=13.2cm]{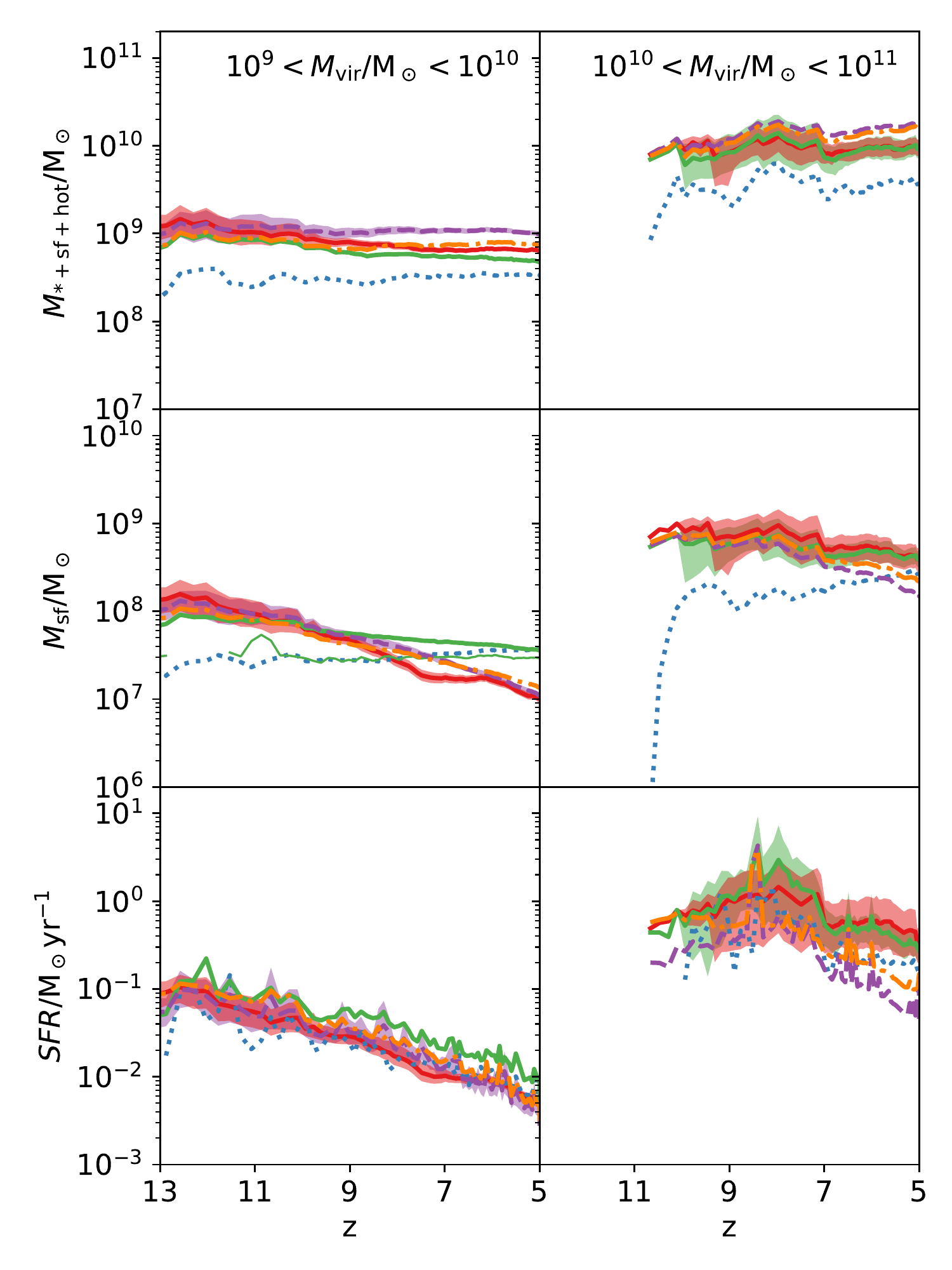}	
	\end{minipage}
	%	\vspace*{-3.8mm}
	\caption{\label{fig:GSMF_histories}\textit{Left panel:} the stellar mass functions at $z\sim11-5$ in the \textit{SAM\_PaperIII}, \textit{SAM\_KS\_limited}, \textit{SAM\_KS\_unlimited}, \textit{SAM\_H2} and {\smaug} \textit{WTHERM} results. Shaded regions represent the $1\sigma$ Poisson uncertainties. Note that at each redshift, only galaxies with $M_\mathrm{vir}>10^9\mathrm{M}_\odot$ are considered and, in order to expand the sample size, we include matched galaxies from 7 consecutive snapshots (${\sim}80$Myr). The resulting galaxy sample size, $N_\mathrm{gals}$, is indicated in the right corner of each subpanel. The two grey regions mark the approximate resolutions of the simulation, which correspond to 10 and 50 stellar particles, respectively. \textit{Right two panels:} the property evolution, including the total baryonic mass, star-forming gas mass and SFR, of galaxies with $10^9{<}M_\mathrm{vir}/\mathrm{M}_\odot{<}10^{10}$ and $10^{10}{<}M_\mathrm{vir}/\mathrm{M}_\odot{<}10^{11}$, respectively. Lines and shaded regions represent the mean and the 95 per cent confidence intervals around the mean using 100000 bootstrap re-samples. In the low-mass panel, the star-forming gas mass of quenched galaxies in the \textit{SAM\_KS\_limited} result is indicated by the thin solid green line (overlapped with the \textit{SAM\_PaperIII} star-forming gas mass).}
\end{figure*}

In order to illustrate this, we use the total-gas-based star formation model (see Section \ref{sec:SAM_SF}) with parameters adopted in \citetalias{Mutch2016a} as an example (see the main SAM parameters in Table \ref{tab:parameters}) and refer to it as \textit{SAM\_PaperIII}. With a short depletion time-scale of the total star-forming gas, the dynamical time-scale for gas transition, and strong supernova feedback, {\meraxes} was able to reproduce the observed stellar mass function at $z=5-7$ in \citetalias{Mutch2016a}. Although we are focusing on lower mass ranges with different dark matter halo merger trees, this combination of parameters can also reproduce the stellar mass function calculated from the {\smaug} hydrodynamic simulation. We show the two stellar mass functions of matched galaxies at $z=11-5$ in the left panel of Fig. \ref{fig:GSMF_histories}. Since only galaxies with $M_\mathrm{vir}>10^{9}\mathrm{M}_\odot$ are included, at each redshift, we combine the result from 7 consecutive snapshots across a time range of ${\sim}80$Myr to obtain adequate samples. We see that the semi-analytic prediction is in agreement with {\smaug} above the resolution limit.

We next show more detailed galaxy property evolution, including the total baryonic mass, star-forming gas mass and SFR, from the two numerical experiments in the right panels of Fig. \ref{fig:GSMF_histories} in two mass ranges. We see that, although the SAM is in agreement with the hydrodynamic simulation on the stellar mass function at a large range of redshifts, they disagree on the evolutionary path of the gas component. We find that the baryonic mass is about $2-5$ times smaller in the SAM compared to the hydrodynamic simulation, suggesting that too much supernova energy has been coupled to the ISM. In addition, the hydrodynamic simulation shows an increasing amount of star-forming gas towards higher redshifts for a given halo mass. This suggests that cooling (or cold-mode accretion) might be more efficient in the early universe. On the other hand, the SAM underestimates the star-forming gas reservoir at higher redshift but predicts a similar SFR. This suggests that the depletion time-scale might be set shorter in the SAM, which happens to result in an agreement with the hydrodynamic result on the stellar mass function.

The relevant free parameters involved in this work are the gas transition time-scale, $t_\mathrm{transition}(\alpha_\mathrm{transition},\beta_\mathrm{transition})$, minimum hydrogen mass for star formation, $m_\mathrm{sf,c}$, depletion time-scales of the total or molecular gas, $t_\mathrm{sf({\htwo})}[\alpha_\mathrm{sf({\htwo})},\beta_\mathrm{sf({\htwo})}]$, coupling efficiency of supernovae energy and the ISM, $\epsilon_\mathrm{energy}(\alpha_\mathrm{energy},\beta_\mathrm{energy})$, and mass loading factor for supernovae heating, $\epsilon_\mathrm{mass}(\epsilon_\mathrm{mass}^\mathrm{max},\alpha_\mathrm{mass},\beta_\mathrm{mass})$. Accounting for normalizations, scaling indices and upper limits, there are 10 free parameters involved. In this case, exploring the full parameter space becomes a challenging task, which is beyond the scope of this work. We are currently constructing an MCMC analysis package for {\meraxes} (Mutch in prep.), and we will apply it to accurately constrain the parameters against observations and the hydrodynamic results in the future. In the following sections, we only consider $\alpha_\mathrm{transition}$, $\beta_\mathrm{transition}$, $\alpha_\mathrm{sf({\htwo})}$, $\beta_\mathrm{sf({\htwo})}$, $\alpha_\mathrm{energy}$ and $\alpha_\mathrm{mass}$ as free parameters when trying to find the best models that can reproduce the hydrodynamic calculation, with the others remaining the same as in \citetalias{Mutch2016a}. We will compare the SAM result with the hydrodynamic simulations with two choices of $m_\mathrm{sf,c}$ and discuss the comparison result in two mass bins ($10^9<M_\mathrm{vir}/\mathrm{M}_\odot<10^{10}$; $10^{10}<M_\mathrm{vir}/\mathrm{M}_\odot<10^{11}$) to explore the impact of the mass scaling indices of supernova feedback ($\beta_\mathrm{energy}$ and $\beta_\mathrm{mass}$). After identifying the fiducial models, we then use them as references and further investigate the parameter space of the SAM.

\subsubsection{Star formation thresholds}\label{sec:sam_ks}

We recalibrate our chosen parameters in order to simultaneously reproduce the evolutions of the stellar mass function of the hydrodynamic simulation, as well as the following three quantities:
\begin{enumerate}
	\item total baryonic mass, which is predominately controlled by supernova ejection;
	\item star-forming gas mass, which is jointly modulated by cooling and supernova heating; and
	\item SFR, which is used to investigate the depletion time-scale of the total or molecular hydrogen gas.
\end{enumerate}
After exploring the parameter space\footnote{We manually explore the parameter space within a plausible range where scaling indices of redshift dependencies (e.g. $\beta_\mathrm{transition}$) are limited to $[-5,0]$.}, we identify a set of parameters (see Table \ref{tab:parameters}) that lead to a better agreement on the property evolution of galaxies with $M_\mathrm{vir}>10^{10}\mathrm{M}_\odot$, which is shown in Fig. \ref{fig:GSMF_histories}. This model is referred to as \textit{SAM\_KS\_limited} and, compared to \textit{SAM\_PaperIII}, this model adopts 
\begin{enumerate}
	\item a smaller coupling efficiency of supernovae energy and the ISM, leading to an increased total baryonic mass;
	\item a longer time-scale of gas transition at $z=5$, which decreases more rapidly towards higher redshifts and results in a better agreement on the star-forming gas mass with the hydrodynamic calculation;
	\item a longer time-scale of gas depletion at $z=5$, which decreases less rapidly towards higher redshifts and is inferred from the star formation efficiency adopted in the hydrodynamic simulation \citepalias{qin2018};
	\item a smaller mass loading factor for supernovae heating to further adjust the star-forming gas mass.
\end{enumerate}
However, in the low-mass range where the virial mass is between $10^9$ and $10^{10}\mathrm{M}_\odot$, this model fails to reproduce the evolutionary path of the star-forming gas reservoir calculated by the hydrodynamic simulation, with a prediction of a flatter $M_\mathrm{sf}-z$ relation. During the experiment, we find that the star-forming gas mass at low redshift ($5<z<8$) does not change by incorporating a larger mass loading factor, $\epsilon_\mathrm{mass}$, which is expected to further suppress the star-forming gas mass through supernova heating. This suggests that the bulk of star-forming gas is stored in quenched galaxies\footnote{In \citetalias{qin2018}, a high barrier to star-forming galaxies was adopted when we studied the total-gas-based star formation prescription in the absence of feedback. However, during the experiment, we found that a more suppressed star formation at higher redshift is required to reproduce the hydrodynamic simulation (i.e. \textit{NOSN\_NOZCOOL\_NoRe}). This is due to the fact that galaxies are more likely to have an insufficient star-forming gas reservoir (i.e. $m_\mathrm{sf}<m_\mathrm{sf,c}$) at lower redshifts due to the adopted high threshold for star formation.} where $m_\mathrm{sf}<m_\mathrm{sf,c}$, the mean star-forming gas mass of which is indicated by the thin solid line in the central panel of Fig. \ref{fig:GSMF_histories}.

According to the total-gas-based star formation prescription (see Section \ref{sec:SAM_SF}), galaxies can only form stars when their gas reservoirs are adequate. This reservoir mass threshold is calculated through
\begin{equation}
m_\mathrm{sf,c} = m_\mathrm{sf,c,0}\times \dfrac{V_\mathrm{vir}}{70 \mathrm{km\ s^{-1}}} \dfrac{r_\mathrm{disc}}{1h^{-1}\mathrm{kpc}},
\end{equation}
with $m_\mathrm{sf,c,0}=1.9\times10^8\mathrm{M}_\odot$ inferred from the KS observations \citep{kennicutt1998global} at the local Universe. In the current DRAGONS series, we instead have adopted a lower critical mass with $m_\mathrm{sf,c,0}=1.4\times10^8\mathrm{M}_\odot$. This is supported by \citet{Henriques2015}, which proposes to reduce the mass threshold of star-forming galaxies to reconcile the issue that previous SAMs have overpredicted the number of quenched galaxies in the low-mass range while these galaxies still possess a significant amount of star-forming gas reservoir. This might explain the evolution of star-forming gas mass of dwarf galaxies predicted by \textit{SAM\_KS\_limited} and suggests that the threshold of star formation needs to be further reduced in these low-mass galaxies.

We next focus on the dwarf galaxies with $10^9\mathrm{M}_\odot<M_\mathrm{vir}<10^{10}\mathrm{M}_\odot$, and recalibrate the model without any thresholds of star formation (i.e. $m_\mathrm{sf,c}=0$). Fig. \ref{fig:GSMF_histories} shows the result of this model, \textit{SAM\_KS\_unlimited}, where gas transition and star formation at lower redshifts as well as the supernovae energy coupling efficiencies are less efficient (the mass loading factor of supernova heating is kept the same as in \textit{SAM\_KS\_limited}). We see that, while \textit{SAM\_KS\_limited} with $m_\mathrm{sf,c}\sim10^8\mathrm{M}_\odot$ is better at reproducing the hydrodynamically simulated high-mass galaxies, the updated model with $m_\mathrm{sf,c}=0$ is more consistent with the hydrodynamic result at the low mass range. This indicates high-redshift less massive galaxies, in general, possess lower thresholds of star formation as well \citep{Henriques2015}.

Note that in \textit{SAM\_KS\_unlimited}, $\beta_{\mathrm{transition}}{=}{-}7.5$ (see equation \ref{eq:cooling}) is crucial to reproducing the evolution of the star-forming gas mass calculated by the hydrodynamic simulation. However, it also leads to unrealistically rapid changes of the gas transition efficiency in the SAM. The time-scale of gas transition from hot to star-forming is close to the dynamical time-scale (55Myr) at $z\sim12$, suggesting strong cold-mode gas inflow at early times \citep{Keres2005,Keres2009,benson2011}. On the other hand, $t_\mathrm{transition}$ increases dramatically towards lower redshifts with $t_\mathrm{transition}$ becoming larger than the Hubble time (1Gyr) at $z\sim8$, leading to greatly suppressed accretion of star-forming gas in low-redshift dwarf galaxies. We note that, with more intense supernova heating to offset it, additional gas can be allowed to transition from hot to star-forming. Therefore, incorporating larger $\epsilon_{\mathrm{mass}}$ at lower redshifts will decrease $t_\mathrm{transition}$ accordingly and allow moderate changes of the transition rate. However, as we will see in Section \ref{sec:supernova_feedback_sam}, supernova heating only plays a secondary role in changing the evolution of star-forming gas of high-redshift dwarf galaxies. We will further discuss this rapidly evolving $t_{\mathrm{transition}}$ in Section \ref{sec:t_transition}. 

In these two \textit{SAM\_KS} models, stars form by consuming the total star-forming gas reservoir. However, due to the unknown time-scale of depleting the total gas (possibly there is no simple universal scaling of $t_\mathrm{sf}$, see \citealt{Duffy2017}), the degeneracy\footnote{Note there are three constraints with four sets of freedoms.} between the processes of cooling and heating exists. For instance, if the \textit{real} depletion time-scale of the total gas reservoir is longer than what we have adopted, cooling would have been underestimated to reproduce the correct SFR. Thus, if we have utilized the true star formation efficiency, we would need to implement 1) more rapid cooling to reproduce the correct amount of stars; 2) a larger mass loading factor to heat the additional star-forming gas that has cooled during the time step; and 3) more efficient coupling between supernova energy and the ISM to provide the energy needed for heating. In the next section, we discuss the $\htwo$-based star formation prescription (see Section \ref{sec:SAM_SF}), which is expected to be a more direct modelling approach and where the relevant time-scale is better understood.

\subsubsection{Star formation from molecular hydrogen}\label{sec:SAM_H2}

\begin{figure*}
	\begin{minipage}{1\textwidth}
		\centering
		\includegraphics[width=.92\textwidth]{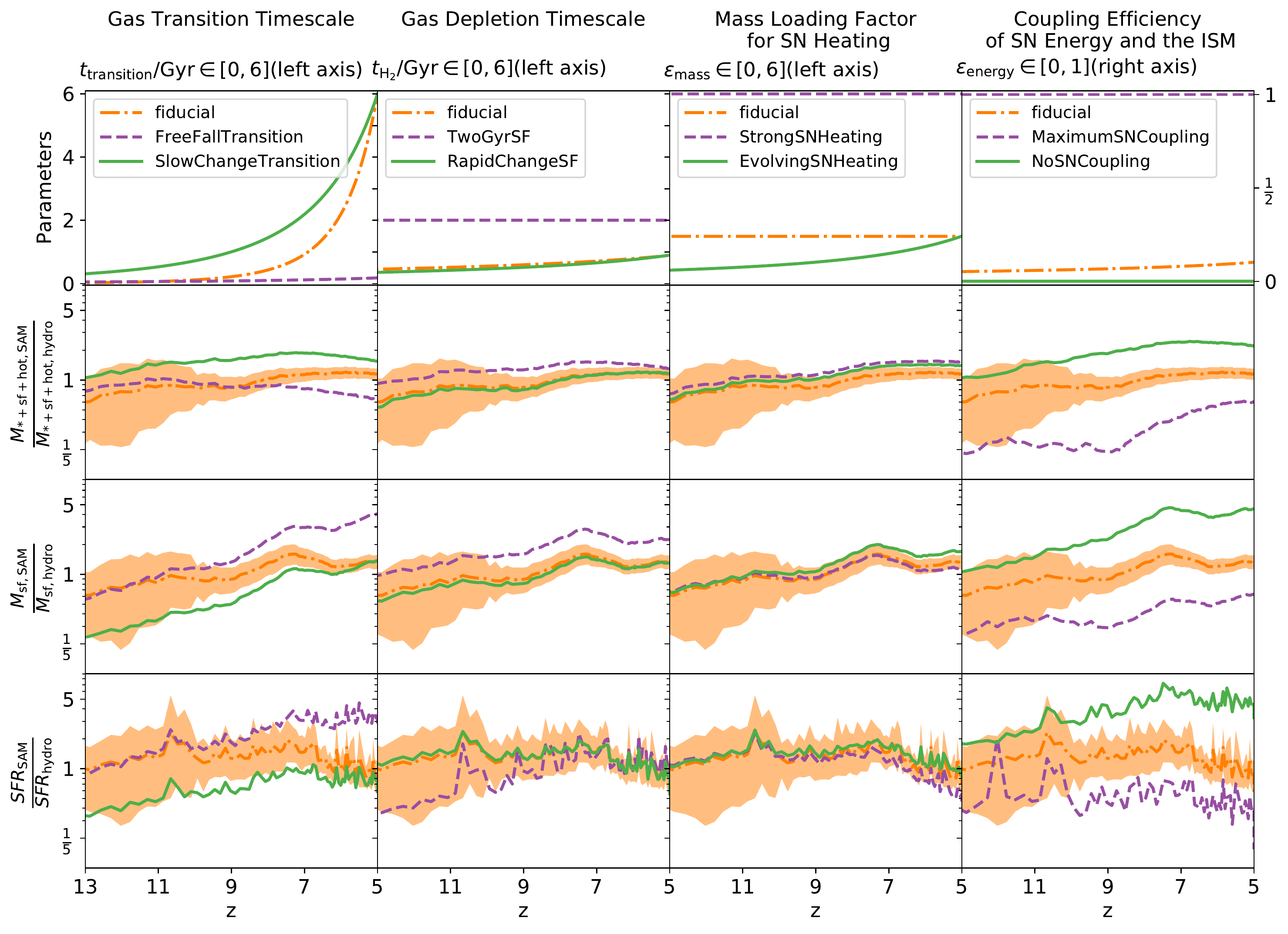}	
	\end{minipage}
		\vspace*{-3mm}
	\caption{\label{fig:varyings_coolingSF} \textit{First row:} the parameter configuration including the gas transition time-scale, gas depletion time-scale, mass loading factor for supernova heating, and coupling efficiency of supernova energy and the ISM, for different models. \textit{Bottom three rows:} the evolution of the ratio of properties calculated by the SAM to the hydrodynamic simulation, including the total baryonic mass, star-forming gas mass and SFR. Only galaxies with $10^9{<}M_\mathrm{vir}/\mathrm{M}_\odot{<}10^{10}$ are considered. Lines and shaded regions (only for the fiducial model) represent the mean and the 95 per cent confidence intervals around the mean using 100000 bootstrap re-samples.}
\end{figure*}

\citet{Duffy2017} investigated the $\htwo$ component of dwarf galaxies using the {\smaug} simulations and found the depletion time-scale of $\htwo$ is $t_{\htwo}\sim0.3\mathrm{Gyr}$, independent of the feedback regime (e.g. \textit{NOSN\_NOZCOOL}, \textit{WTHERM}). They also discussed the mass and redshift dependencies when applying SAMs with $\htwo$-based star formation laws and proposed\footnote{We ignore the weak dependency on stellar mass and use $10^{7.2}\mathrm{M}_\odot$, which is approximately the average stellar mass of the final sample, as a representative in this work.} $t_{\htwo}\sim0.9\mathrm{Gyr}\left[\left(1{+}z\right){/}6\right]^{-1.1}$, the extrapolation of which agrees with the previous findings at the local Universe \citep{Leroy:2008jk}. We note that the scaling index of $t_\mathrm{\htwo}$ was motivated from the KS law with an assumption that galactic discs are self-gravitating and follow exponential surface profiles. The latter might need revising for high-redshift dwarf galaxies. In the early universe, galaxies tend to possess larger velocity dispersions and both mergers and cold-mode accretion are significant. These all indicate that high-redshift galaxies might have thickened discs \citep{Moster2012,Newman2012,Price2015}. In this section, we adopt $t_{\htwo}\sim0.9\mathrm{Gyr}\left[\left({1+z}\right)/6\right]^{-0.8}$ instead with the scaling inferred from an isothermal profile, and then calibrate cooling and supernova feedback efficiencies to reproduce the dwarf galaxy properties from {\smaug}. We will further discuss the semi-analytic prediction when varying $t_{\htwo}$ in Section \ref{sec:t_h2}.

The result is shown in Fig. \ref{fig:GSMF_histories}. We see that without any degeneracies\footnote{Note there are three constraints with three sets of freedoms.}, the $\htwo$-based model can still reproduce the hydrodynamic calculation of the properties of dwarf galaxies as well as the cosmic evolution of the stellar mass function. Compared to the \textit{SAM\_KS\_unlimited} result, \textit{SAM\_H2} agrees better with {\smaug} on the evolution of the total baryonic mass of dwarf galaxies, and the calculation of massive galaxies. However, it still overestimates the total baryonic mass and underestimates the mass of star-forming gas and SFR of massive galaxies, suggesting that these galaxies might possess different cooling and supernova feedback efficiencies or shorter depletion time-scale of $\htwo$ compared to less massive galaxies \citep{Duffy2017}.

The success of our SAM with the $\htwo$-based star formation prescription and a fixed $\htwo$ depletion time-scale is encouraging. It indicates that the accretion--cooling--depletion--heating--and--ejection pathway of gas is still representative for dwarf galaxy formation at high redshift in terms of predicting the gas and stellar properties of the hydrodynamic simulation. We next use the $\htwo$-based SAM as an example and illustrate the impact of changing the relevant parameters with comparisons to the fiducial \textit{SAM\_H2} model. In practice, we change one set of parameters at each time with the other parameters remaining the same as the fiducial model. We show the result of the property evolution in Fig. \ref{fig:varyings_coolingSF} in terms of the ratio of the properties calculated by the SAM to the hydrodynamic simulation.

\subsubsection{Gas transition time-scale}\label{sec:t_transition}

The gas transition time-scale of the fiducial \textit{SAM\_H2} model is $t_\mathrm{transition}=6\left[\left({1+z}\right)/6\right]^{-6.5}\mathrm{Gyr}$, which results in a significantly larger value at low redshift. In the first column of Fig. \ref{fig:varyings_coolingSF}, we show this scaling, as well as the result of assigning $t_\mathrm{transition}$ with the free-fall time-scale (\textit{FreeFallTransition}), which is a common assumption adopted in the literature for the rapid cooling regime, and a time-scale that evolves slower towards higher redshifts (\textit{SlowChangeTransition}) compared to the fiducial model. We see that adopting a shorter transition time-scale results in the star-forming gas reservoir receiving more efficient replenishment. In the case of unchanged gas depletion time-scale, SFR increases. Consequently, more energy gets ejected from supernova explosions, leading to more suppressed total baryonic masses given that the energy coupling efficiency and mass loading factor for heating do not change. From the first column of Fig. \ref{fig:varyings_coolingSF}, we see that without changing other parameters, a significantly evolving transition time-scale is required to reproduce the rapidly decreasing star-forming gas mass at lower redshifts as predicted by the hydrodynamic simulation.

We note again that varying $t_\mathrm{transition}$ from the dynamical time-scale was proposed to compensate for the overestimated collapse rate attained by assuming SIS profiles for hot gas and the underestimation due to large transition radii between hot and star-forming gas \citepalias{qin2018}. In the presence of feedback, the star-forming disc shrinks, in particular at lower redshifts (see Appendix \ref{sec:dis_gas_transition}). Correspondingly, the overestimation of $t_\mathrm{transition}$ (overestimated $L_\mathrm{inflow}$ in equation \ref{eq:t_transition}) becomes insignificant -- gas indeed needs to collapse into the central region to become star-forming gas. Therefore, we need a much longer transition time-scale at low redshift to take into account that hot gas is less dense at the inner regions compared to the SIS profile -- only a small amount of gas can collapse into the centre within the dynamical time-scale (overestimated $\dot{m}_\mathrm{hot->sf}$ in equation \ref{eq:t_transition}). We will investigate below whether the other free parameters can have a significant impact to the star-forming gas reservoir evolution, so that the steep gradient of $m_\mathrm{sf}(z)$ in Fig. \ref{fig:GSMF_histories} can be reproduced without the implementation of a rapidly-evolving redshift dependency in $t_\mathrm{transition}$.

\subsubsection{Gas depletion time-scale}\label{sec:t_h2}
The $\htwo$ depletion time-scale is better constrained than that of the total gas. However, observational results still possess large variance even in the local Universe, from a half to a few Gyr \citep{Leroy:2008jk,Bigiel:2011hp,Saintonge:2011ey,Tacconi:2013cz}. Therefore, we use the redshift-dependent molecular hydrogen depletion time-scale proposed in \citet{Duffy2017}, which rapidly decreases at higher redshift (see Section \ref{sec:SAM_SF}). For the fiducial model, we use $t_\mathrm{\htwo}=0.9\mathrm{Gyr}\times\left[\left({1+z}\right)/6\right]^{-0.8}$, with a less steeply evolving redshift dependency due to thicker discs at high redshift (see Section \ref{sec:SAM_H2}). In the second column of Fig. \ref{fig:varyings_coolingSF}, we show the property evolution of using the time-scales proposed by \citet{Duffy2017} (\textit{RapidChangeSF}) as well as a constant value (\textit{TwoGyrSF}), $t_\mathrm{\htwo}=2 \mathrm{Gyr}$, which is commonly adopted for SAMs in the literature (e.g. \citealt{Lagos2011}). We see that by increasing the time-scale of converting hydrogen into stars, star formation quenches, leading to weaker supernova ejection and heating. With the current configuration of parameters, we see that $t_\mathrm{\htwo}=2\mathrm{Gyr}$ significantly underestimates star formation at high redshift in agreement with \citet{Duffy2017} and the star-forming gas evolution gradient is not expected to change significantly by varying $\beta_\mathrm{\htwo}$.

\subsubsection{Supernova feedback parameters}\label{sec:supernova_feedback}
\begin{figure}
	\centering
	\includegraphics[width=.85\columnwidth]{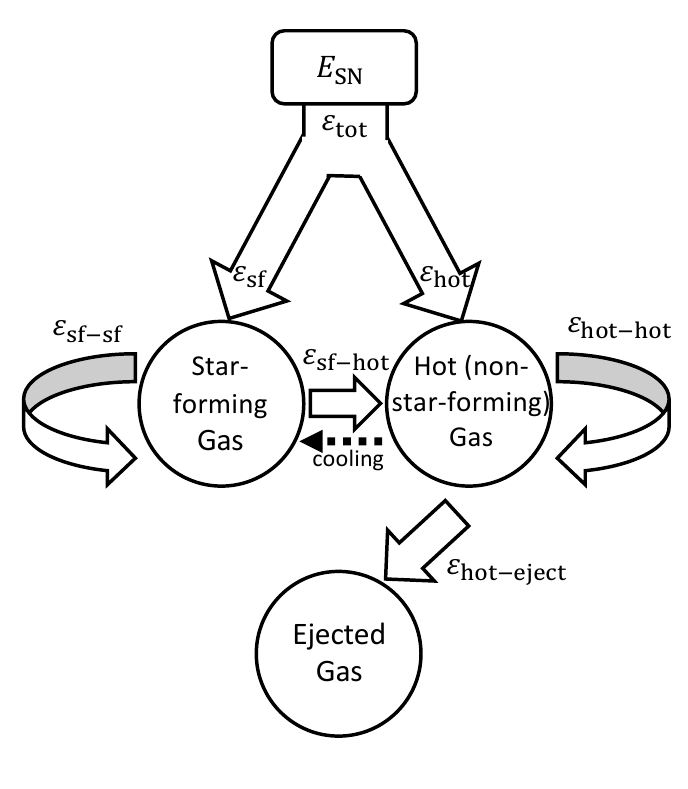}\vspace*{-2mm}
	\caption{\label{fig:fSN}Energy flow of supernova feedback. $\epsilon_\mathrm{tot}$, $\epsilon_\mathrm{sf}$, $\epsilon_\mathrm{hot}$, $\epsilon_\mathrm{sf-sf}$, $\epsilon_\mathrm{sf-hot}$, $\epsilon_\mathrm{hot-hot}$ and $\epsilon_\mathrm{hot-eject}$ represent fractions of supernova energy ($E_\mathrm{SN}$) that are 1) coupled to the ISM; 2) distributed to the star-forming gas; 3) distributed to the hot gas; 4) used to increase the thermal energy of the star-forming gas reservoir; 5) used to heat the star-forming gas to hot; 6) used to increase the hot gas thermal energy; and 7) used to eject the hot gas from the host. Note that $\epsilon_\mathrm{tot}=\epsilon_\mathrm{sf}+\epsilon_\mathrm{hot}$, $\epsilon_\mathrm{sf}=\epsilon_\mathrm{sf-sf}+\epsilon_\mathrm{sf-hot}$ and $\epsilon_\mathrm{hot}=\epsilon_\mathrm{hot-hot}+\epsilon_\mathrm{hot-eject}$. The dotted line indicates the process of radiative cooling.}
\end{figure}
Supernova explosions increase the thermal energy of the ISM and expel baryons in dwarf galaxies. However, since the relevant region cannot be resolved in cosmological simulations, subgrid physics with free parameters are adopted by both hydrodynamic and semi-analytic modelling approaches. These parameters (see Section \ref{sec:models}) represent the coupling efficiency between supernova energy and the ISM (i.e. $\epsilon_\mathrm{energy}$ and $f_\mathrm{th}$) or modulate heating and ejection by supernova feedback (i.e. $\epsilon_\mathrm{mass}$ and $\Delta T$). We use the energy flow illustrated in Fig. \ref{fig:fSN} to facilitate the following discussions of supernova feedback, where
\begin{enumerate}
	\item $\epsilon_\mathrm{tot}$ represents the fraction of the energy ejected by supernovae ($E_\mathrm{SN}$) that is coupled to the ISM;
	\item $\epsilon_\mathrm{sf}$ and $\epsilon_\mathrm{hot}$ represent the fractions of energy that are distributed to the star-forming and hot gas, respectively, with $\epsilon_\mathrm{tot}=\epsilon_\mathrm{sf}+\epsilon_\mathrm{hot}$;
	\item $\epsilon_\mathrm{sf-sf}$ and $\epsilon_\mathrm{sf-hot}$ represent the fractions of energy that are used to increase the thermal energy of the star-forming gas reservoir and to heat the star-forming gas to hot, respectively, with $\epsilon_\mathrm{sf}=\epsilon_\mathrm{sf-sf}+\epsilon_\mathrm{sf-hot}$; and
	\item $\epsilon_\mathrm{hot-hot}$ and $\epsilon_\mathrm{hot-eject}$ represent the fractions of energy that are used to increase the thermal energy of the hot gas reservoir and to eject the hot gas from the host, respectively, with $\epsilon_\mathrm{hot}=\epsilon_\mathrm{hot-hot}+\epsilon_\mathrm{hot-eject}$.
\end{enumerate}

\textbf{Energy-ISM coupling efficiency:} The fraction of supernova energy that contributes to feedback is $f_\mathrm{th}$ and $\epsilon_\mathrm{energy}$ in {\smaug} and {\meraxes}, respectively. Since $f_\mathrm{th}$ is chosen to be unity, with all the supernova energy being coupled to the ISM, one might expect $\epsilon_\mathrm{energy}=1$ as well. However, because SAMs ignore the thermal energy of the star-forming gas and assume the temperature of hot gas does not change during one time step, $\epsilon_\mathrm{sf-sf}$ and $\epsilon_\mathrm{hot-hot}$ are zero. This means that, despite all supernova energy being coupled to the ISM in the hydrodynamic simulation (i.e. $f_\mathrm{th}\equiv\epsilon_\mathrm{tot}=1$), only a fraction of it contributes to feedback in the SAM (some of the energy might also be lost in the ejected gas reservoir due to the untraced thermal energy of the ejected gas). Therefore, $\epsilon_\mathrm{energy}\equiv \epsilon_\mathrm{sf-hot}+\epsilon_\mathrm{hot-eject}<f_\mathrm{th}=1$.

\textbf{Mass loading factor:} How much of the supernova energy is coupled to the ISM and used to heat gas ($\Delta m_\mathrm{sf}$) is governed by free parameters describing the mass loading factor ($\epsilon_\mathrm{mass,sam}$) in the SAM while in the hydrodynamic simulation, it is the increment of gas temperature ($\Delta T$) that determines the number of gas particles that are affected. This indicates that, in the hydrodynamic simulation,
\begin{equation}
\begin{split}
\dfrac{3}{2}\dfrac{\Delta m_\mathrm{sf}}{\mu m_\mathrm{p}}k_\mathrm{B}\Delta T& = 10^{51}\mathrm{erg} \times f_\mathrm{th}\eta_\mathrm{SNII}\Delta m_*,
\end{split}
\end{equation}
where $\mu m_\mathrm{p}$ is the average particle mass of fully ionized gas. With a large temperature increment of $\Delta T=10^{7.5}\mathrm{K}$, the mean number of instantaneously heated nearby gas particles per stellar baryon is $\epsilon_{\mathrm{mass,hydro,ins}}\equiv\Delta m_\mathrm{sf}/m_*\sim1.34$ \citep{DallaVecchia2012}. This small mass loading factor places the heated gas in the Bremsstrahlung cooling regime, achieving an efficient supernova feedback mechanism through heating. Moreover, due to the increased pressure from the thermal feedback, gas particles within high overdensities tend to move outwards in a wind, perpendicularly to the star-forming disc plane as a result of the density gradient. These particles are usually referred to as the wind particles, some of which might eventually escape from the gravitational potential of the system. As the wind particles travel, they further increase the thermal energy of the nearby gas particles along the path, leading to a much larger effective mass loading factor over a long period of time such that $\epsilon_{\mathrm{mass,hydro,eff}}\gg 1.34$. Since the SAM captures the average property over 11Myr, one might expect $\epsilon_{\mathrm{mass,sam}}\gg1.34$ as well. However we have also shown that in hydrodynamically simulated dwarf galaxies, gas particles need not be fully virialized to become non-star-forming gas (i.e. $V_\mathrm{hot,hydro} <V_\mathrm{vir}$; \citetalias{qin2018}) while on the other hand, SAMs ignore the thermal energy of the star-forming gas ($V_\mathrm{sf,sam}=0$) and assume the non-star-forming hot gas shares the virial temperature of host halo ($V_\mathrm{hot,sam}=V_\mathrm{vir}$). Therefore, following the supernova energy used for heating in the SAM and hydrodynamic simulation ($E_\mathrm{reheat,sam} = E_\mathrm{reheat,hydro}$), we see that
\begin{equation}
\begin{split}
\epsilon_{\mathrm{mass,sam}} &\equiv \dfrac{E_\mathrm{reheat,sam}}{\dfrac{1}{2} \Delta m_\mathrm{sf} \left(V_\mathrm{hot,sam}^2{-}V_\mathrm{sf,sam}^2\right)} {=} \dfrac{E_\mathrm{reheat,hydro}}{\dfrac{1}{2} \Delta m_\mathrm{sf} V_\mathrm{vir}^2}\\
&\equiv\dfrac{\dfrac{1}{2} \epsilon_{\mathrm{mass,hydro,eff}} \Delta m_\mathrm{sf} \left<V_\mathrm{hot,hydro}^2 {-} V_\mathrm{sf,hydro}^2\right>}{\dfrac{1}{2} \Delta m_\mathrm{sf} V_\mathrm{vir}^2}\\
&<\epsilon_{\mathrm{mass,hydro,eff}},
\end{split}
\end{equation}
where $0.5V_\mathrm{hot,sam(hydro)}^2$ and $0.5V_\mathrm{sf,sam(hydro)}^2$ represent the specific thermal energy of the hot and star-forming gas reservoirs in the SAM (or particles in the hydrodynamic simulation) while $\left<\right>$ indicate the property is averaged over all reheated gas particles and a long period of time in the hydrodynamic simulation.

\subsubsection{Supernova feedback in the SAM}\label{sec:supernova_feedback_sam}
Without properly tracking the thermal energy of varied gas reservoirs in the SAM, it is challenging to determine the energy-ISM coupling efficiency and mass loading factor. In this work, against the hydrodynamic result of \textit{WTHERM}, we have calibrated our fiducial SAM and adopted $\epsilon_{\mathrm{mass}}{=}1.5$ and $\epsilon_{\mathrm{energy}}{=}0.03\left(\dfrac{1}{2}+V_\mathrm{max,70}^{-2}\right)$ for \textit{SAM\_H2} (see Table \ref{tab:parameters}). In the last two columns of Fig. \ref{fig:varyings_coolingSF}, we show four varying models of supernova feedback with 1) stronger supernova heating (\textit{StrongSNHeating}); 2) an evolving mass loading factor (\textit{EvolvingSNHeating}) with $\beta_\mathrm{mass}=3.5$ following \citet{guo2011dwarf}; 3) no coupling between supernova energy and the ISM (\textit{NoSNCoupling}); and 4) all supernova energy contributing feedback (\textit{MaximumSNCoupling}) in terms of converting star-forming gas to hot and ejecting hot gas from galaxies ($\epsilon_\mathrm{sf-hot}{+}\epsilon_\mathrm{hot-eject}{=}1$, see the illustration in Fig. \ref{fig:fSN}). Note that $V_\mathrm{max}\sim V_\mathrm{vir}=41\left[\left({1+z}\right)/6\right]^{0.5}\mathrm{km/s}$ for haloes with $M_\mathrm{vir}\sim10^{9.5}\mathrm{M}_\odot$. Therefore, the velocity dependencies (see equation \ref{eq:epsilon_energy}) that were proposed to supernova feedback \citep{guo2011dwarf} represent a cosmic evolutionary path for a given halo mass.
% For instance,
%\begin{equation}
%\begin{split}
%\epsilon_\mathrm{energy}&=\alpha_\mathrm{energy}\left(\dfrac{1}{2}+V_\mathrm{max,70}^{-\beta_\mathrm{energy}}\right)\\
%&\sim\alpha_\mathrm{energy}\left[\dfrac{1}{2}+\left(\dfrac{1+z}{10}\right)^{-0.5\beta_\mathrm{energy}}\right].
%\end{split}
%\end{equation}

We see that when the mass loading factor is fixed, more coupled energy to the ISM leads to stronger suppression of the total baryonic mass, which subsequently decreases the mass of the star-forming disc and quenches star formation. From the \textit{MaximumSNCoupling} model, we see that with all supernova energy used to convert star-forming gas to hot and eject hot gas from the galaxy, the total baryonic mass and star-forming gas become significantly suppressed. On the other hand, when the supernova energy coupling efficiency is fixed, less heating leads to a larger reservoir of star-forming gas and enhanced star formation. Consequently, more supernova energy is coupled to the ISM. With less energy used for heating, more mass in the hot gas reservoir gets ejected. Depending on the increased amount of star-forming gas and stellar mass as well as the decreased hot gas mass, the total baryonic mass varies slightly. For instance, from the \textit{EvolvingSNHeating}, to fiducial and \textit{StrongSNHeating} model, mass loading factor increases, leading to decreased SFR and $M_\mathrm{sf}$. However, both \textit{EvolvingSNHeating} and \textit{StrongSNHeating} predict more baryonic mass compared to the fiducial model, with more star-forming gas and stars formed in the former while more hot gas is retained in the latter model. In addition, the property evolution does not change significantly between these three models. Therefore, we do not expect that, by changing the heating efficiency of supernovae, the issue of incorporating a rapidly evolving gas transition time-scale (see Section \ref{sec:t_transition}) can be resolved. 

\section{Conclusions}\label{sec:conclusion}
Following \citet[][\citetalias{qin2018}]{qin2018}, we further investigate the semi-analytic modelling prescriptions of galaxy formation that are commonly adopted in the literature. In this work, we include supernova feedback and homogeneous reionization background in both the {\meraxes} SAM \citep{Mutch2016a} and \textit{Smaug} high-resolution hydrodynamic simulation \citep{duffy2014low}, and make comparisons between the stellar and gas reservoirs predicted by these two methods. We focus on galaxies with $10^9\mathrm{M}_\odot<M_\mathrm{vir}\lesssim 10^{11}\mathrm{M}_\odot$. With the modifications previously proposed in \citetalias{qin2018} including adjustments to halo masses from the merger trees, suppression of baryon fractions accounting for hydrostatic pressures, and the modulation of time-scales for the transition of gas from hot to star-forming ($t_\mathrm{transition}$) and from star-forming to stars (depletion time-scale, $t_\mathrm{sf({\htwo})}$), we find that the current SAM is able to reproduce the hydrodynamic calculation of the cosmic evolution of galaxies with $M_\mathrm{vir}>10^{10}\mathrm{M}_\odot$ at high redshift. This includes the stellar mass function, total baryonic mass, star-forming gas mass and SFR between $z=5-11$. However, in less massive galaxies ($10^9\mathrm{M}_\odot<M_\mathrm{vir}< 10^{10}\mathrm{M}_\odot$) with SFR calculated using the total star-forming gas, we identify a significant amount of star-forming gas stored in quenched galaxies due to the imposed mass threshold of star formation. After reducing the threshold, the SAM successfully mimics the evolution of dwarf galaxies in the hydrodynamic simulation. 

We also investigate a second star formation prescription, which splits the star-forming gas disc into molecular and atomic hydrogen and forms stars from molecules \citep{Lagos2011}. Fixing the depletion time-scale of $\htwo$ inferred from a previous study of the {\smaug} hydrodynamic simulation \citep{Duffy2017}, we find that, with only calibrations of the gas transition rate and supernova efficiencies, the SAM can also reproduce the dwarf galaxy properties calculated by the hydrodynamic simulation. In addition, we find that when reionization and supernova feedback are included, dwarf galaxies tend to accrete a significant amount of star-forming gas at early times ($z>10$), which quickly becomes suppressed towards lower redshifts. Future work needs to take this into account and incorporate modelling of cold-mode accretion to study dwarf galaxies in the early universe.

\section*{Acknowledgements}
This research was supported by the Victorian Life Sciences Computation Initiative, grant ref. UOM0005, on its Peak Computing Facility hosted at the University of Melbourne, an initiative of the Victorian Government, Australia. Part of this work was performed on the gSTAR national facility at Swinburne University of Technology. gSTAR is funded by Swinburne and the Australian Governments Education Investment Fund. This research was conducted by the Australian Research Council Centre of Excellence for All Sky Astrophysics in 3 Dimensions (ASTRO 3D), through project number CE170100013. This work was supported by the Flagship Allocation Scheme of the NCI National Facility at the ANU, generous allocations of time through the iVEC Partner Share and Australian Supercomputer Time Allocation Committee. YQ acknowledges support from the Albert Shimmins Fund. AM acknowledges support from the European Research Council under the European Union's Horizon 2020 research and innovation program (Grant No. 638809 -- AIDA).
\bibliographystyle{\dir mn2e}
\bibliography{reference}

\appendix
\section{Modifications of halo masses and baryon fractions}\label{sec:halo mass}

\begin{figure*}
	\includegraphics[width=.95\textwidth]{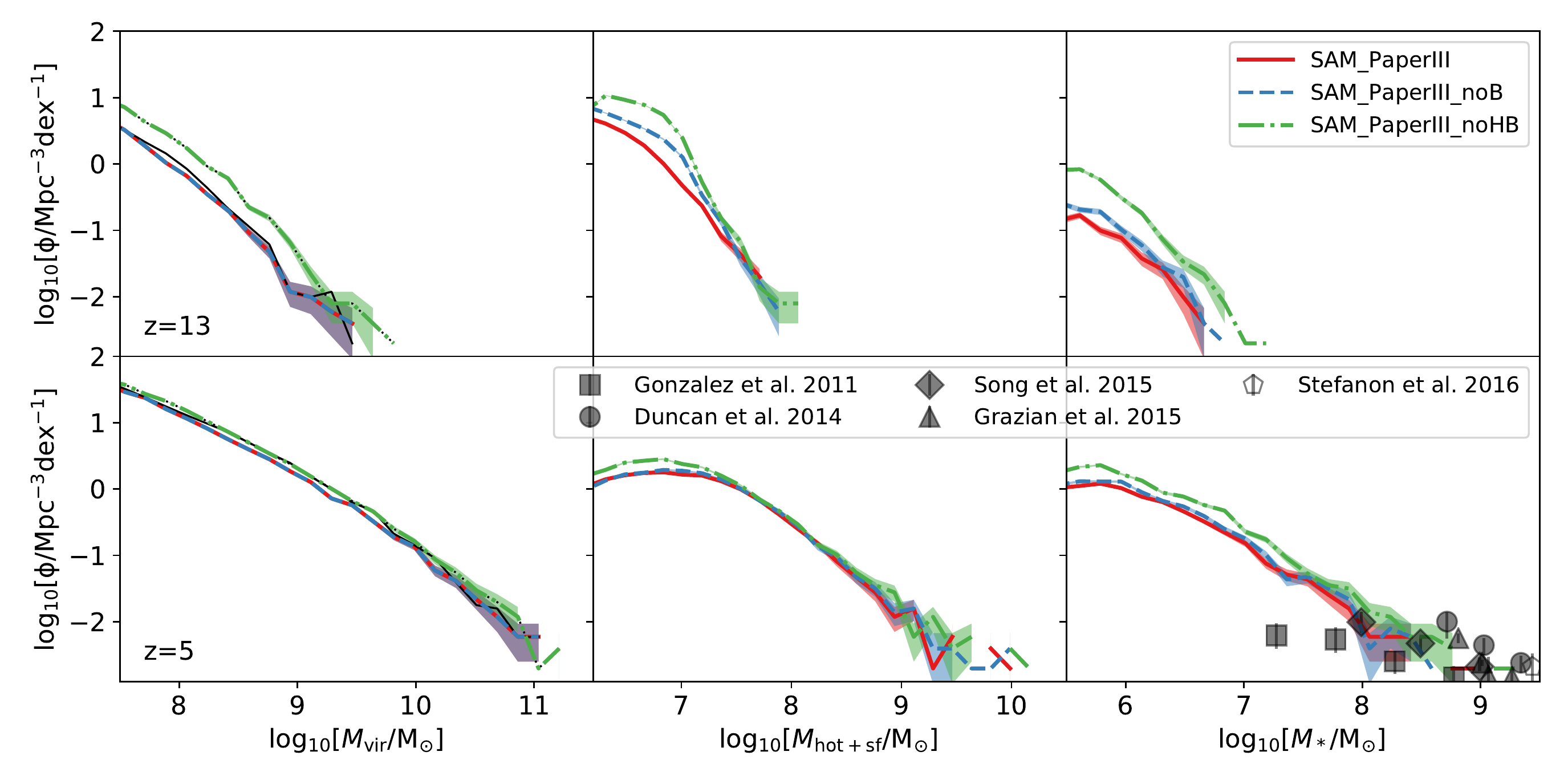}%\vspace*{-2mm}
	\caption{\label{fig:modifiers}Comparisons of the virial mass, gas mass and stellar mass functions (from left to right) at $z=5$ and 12 between three {\meraxes} \textit{WTHERM} runs with fixed parameters adopted by \citetalias{Mutch2016a} but different implementations of modifiers: 1) no modifiers (\textit{SAM\_PaperIII\_noHB}); 2) halo mass modifier only (\textit{SAM\_PaperIII\_noB}); and 3) two modifiers including halo mass and baryon fraction (\textit{SAM\_PaperIII}). Shaded regions represent the $1\sigma$ Poisson uncertainties. For comparison, the halo mass functions predicted by \textit{WTHERM} {\smaug} hydrodynamic simulation and the \textit{N}-body simulation are indicated by the thin black solid and dotted lines, respectively.}
\end{figure*}

We show the impact to the semi-analytic calculation, in the presence of reionization and supernova feedback, of incorporating the halo mass and baryon fraction modifiers \citep{Qin2017a,qin2018}, which correspond to the slower evolution of haloes and less efficient gas accretion due to hydrostatic pressure. We apply {\meraxes} with the total-gas-based star formation law (see Section \ref{sec:SAM_SF}) and the same parameters adopted in \citetalias{Mutch2016a} (\textit{SAM\_PaperIII}) but without the baryon fraction modifier (\textit{SAM\_PaperIII\_noB}) and without any modifiers (\textit{SAM\_PaperIII\_noHB}). We show the halo mass, gas mass and stellar mass functions at $z=5$ and 12 of the three {\meraxes} results and the halo mass functions predicted by the {\smaug} simulations in Fig. \ref{fig:modifiers}. 

We see that without the halo mass modifier, the halo mass function is overestimated compared to the hydrodynamic result at high redshift, which consequently increases the mass function of gas and stars. In addition, further excluding the baryon fraction modifier increases the amount of gas accreted by the host halo and subsequently causes more stars to form. However, we see that the modifications have an insignificant impact to the stellar mass function in the current observable range, which requires deeper surveys with upcoming space programs such as JWST\footnote{\url{https://jwst.nasa.gov/}}.

\section{The build-up of star-forming gas}\label{sec:dis_gas_transition}

\begin{figure*}
	\begin{minipage}{\textwidth}
		\includegraphics[width=\textwidth]{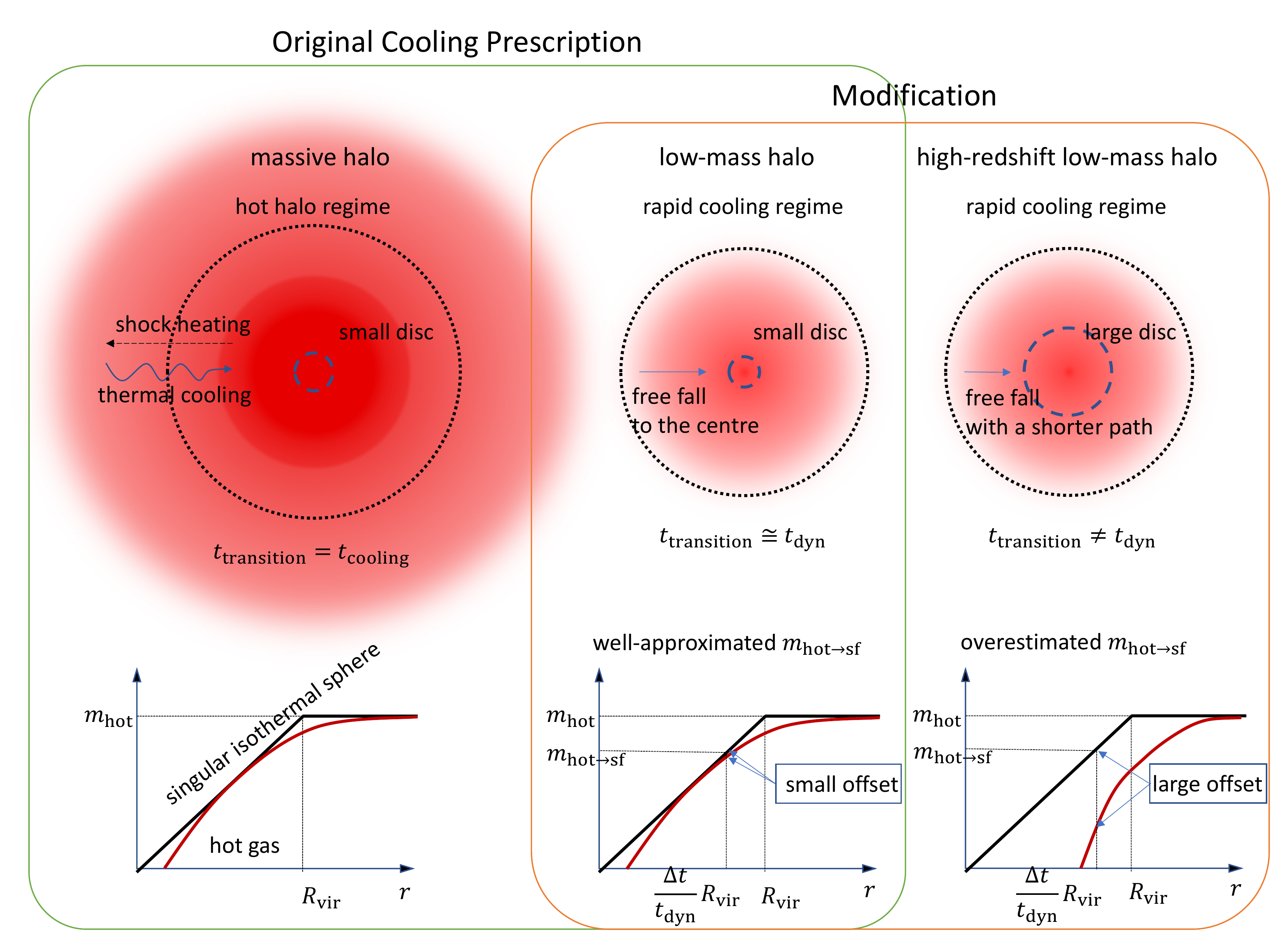}
	\end{minipage}
	\caption{\label{fig:cooling_prescription}Modifications to the semi-analytic cooling prescription. \textit{Top panels:} the contour demonstrate the gas density with the outermost regions illustrating the size of the hot halo, $R_\mathrm{vir}$, while the dotted line indicates the cooling radius, at which the cooling time is equal to the halo dynamical time. \textit{Bottom panels:} the illustration of the hot gas density profile compared to the SIS profile. In the original cooling prescription, hot gas is assumed to share the halo virial temperature and follows the SIS profile. Gas of massive haloes usually remains in thermal equilibrium due to shock heating, and the transition time-scale of gas from hot to star-forming depends on the thermal cooling time-scale, $t_\mathrm{transition}=t_\mathrm{cool}$. On the other hand, gas of less massive haloes do not experience significant heating from shocks, and gas falls onto the central disc at the dynamical time-scale, $t_\mathrm{transition}\approx t_\mathrm{dyn}$. The modification proposed to the rapid cooling regime is that during the rapid cooling regime, $t_\mathrm{transition} \ne t_\mathrm{dyn}$.}
\end{figure*}

\citet[][\citetalias{qin2018}]{qin2018} shows that in the absence of feedback, the majority of dwarf galaxies in the hydrodynamic simulation accrete gas particles with temperatures around a few of $10^4\mathrm{K}$, which is much lower than their halo virial temperatures. This represents a cold-mode accretion of the infalling gas \citep{Keres2005,Keres2009}, which in the SAM is currently modelled through the cooling prescription of the \textit{rapid cooling regime} proposed by \citet{white1991} -- gas accreted in hot mode shares the halo virial temperature due to shock heating and cools rapidly within the dynamical time-scale, $t_\mathrm{dyn}$. The infalling hot gas ($m_\mathrm{hot}$) is also assumed to follow the singular isothermal sphere (SIS) profile
\begin{equation}
\mathscr{M}_\mathrm{SIS}\left(r\right) = m_\mathrm{hot}\times \min\left(\dfrac{r}{R_\mathrm{vir}},1\right).
\end{equation} 
To ease demonstration in this paper, we term the time-scale of gas being transited from hot reservoir to the star-forming disc as a transition time-scale
\begin{equation}\label{eq:t_transition}
t_\mathrm{transition}\equiv\dfrac{m_\mathrm{hot->sf}}{\dot{m}_\mathrm{hot->sf}}\equiv\dfrac{L_\mathrm{inflow}}{{V}_\mathrm{inflow}},
\end{equation}
where $m_\mathrm{hot->sf}$ and $\dot{m}_\mathrm{hot->sf}$ represent the mass and mass rate of the transition, while $L_\mathrm{inflow}$ and $V_\mathrm{inflow}$ are the distance and velocity of the corresponding gas inflow, respectively. We illustrate the cooling prescription in Fig. \ref{fig:cooling_prescription}.
Most massive haloes are able to create shocks and heat the infalling gas, resulting in hydrostatic equilibrium. In this case, which is termed the \textit{hot halo regime}, the time-scale of hot gas transitioning to star-forming is determined by the thermal cooling time-scale, $t_\mathrm{transition}=t_\mathrm{cool}$. However, it is difficult to generate shock heating in less massive systems \citep{Birnboim2003,Cattaneo2017}, leaving little support to prevent gas from infalling onto the central disc, and cooling becomes rapid. In this \textit{rapid cooling regime}, the prescription assumes the star-forming gas disc is relatively small and such a process happens as free-fall. These mean that the gas at the virial radius needs to travel through a distance of $L_\mathrm{inflow}\approx R_\mathrm{vir}$ with $V_\mathrm{inflow}=V_\mathrm{vir}$, leading to $t_\mathrm{transition}\approx t_\mathrm{dyn}$. 

In making comparisons of the gas reservoir calculated by the SAM and hydrodynamic simulation with reionization and supernova feedback isolated in \citetalias{qin2018}, we found $t_\mathrm{transition}=t_\mathrm{dyn}$ becomes less accurate when applying the rapid cooling prescription to high-redshift dwarf galaxy modelling. This is due to the aforementioned two assumptions which lead to over- and under-estimations of the gas transited from hot to star-forming, respectively.
\begin{enumerate}
	\item Assuming the SIS profile of the accreted mass overestimates the gas density in the inner regions (see the illustration in the bottom panels of Fig. \ref{fig:cooling_prescription}). Subsequently, for a given time step of $\Delta t<t_\mathrm{dyn}$, $m_\mathrm{hot->sf}=m_\mathrm{hot}\dfrac{\Delta t}{t_\mathrm{dyn}}=\mathscr{M}_\mathrm{SIS}\left(r=\dfrac{\Delta t}{t_\mathrm{dyn}}R_\mathrm{vir}\right)$ is overestimated;
	\item star-forming gas particles of dwarf galaxies (in the hydrodynamic simulation) possess larger extensions and can be found as far as the virial radius. This means that assuming gas can only transfer from non-star-forming hot gas to star-forming when it reaches the galaxy centre introduces a longer inflow path ($L_\mathrm{inflow}$) and hence leads to an overestimated transition time-scale ($t_\mathrm{transition}$; see equation \ref{eq:t_transition}). In this case, for a given time step of $\Delta t$, $m_\mathrm{hot->sf}=m_\mathrm{hot}\dfrac{\Delta t}{t_\mathrm{transition}}$ is underestimated instead.
\end{enumerate}

\begin{figure}
	\begin{minipage}{\columnwidth}
		\centering
		\includegraphics[width=\textwidth]{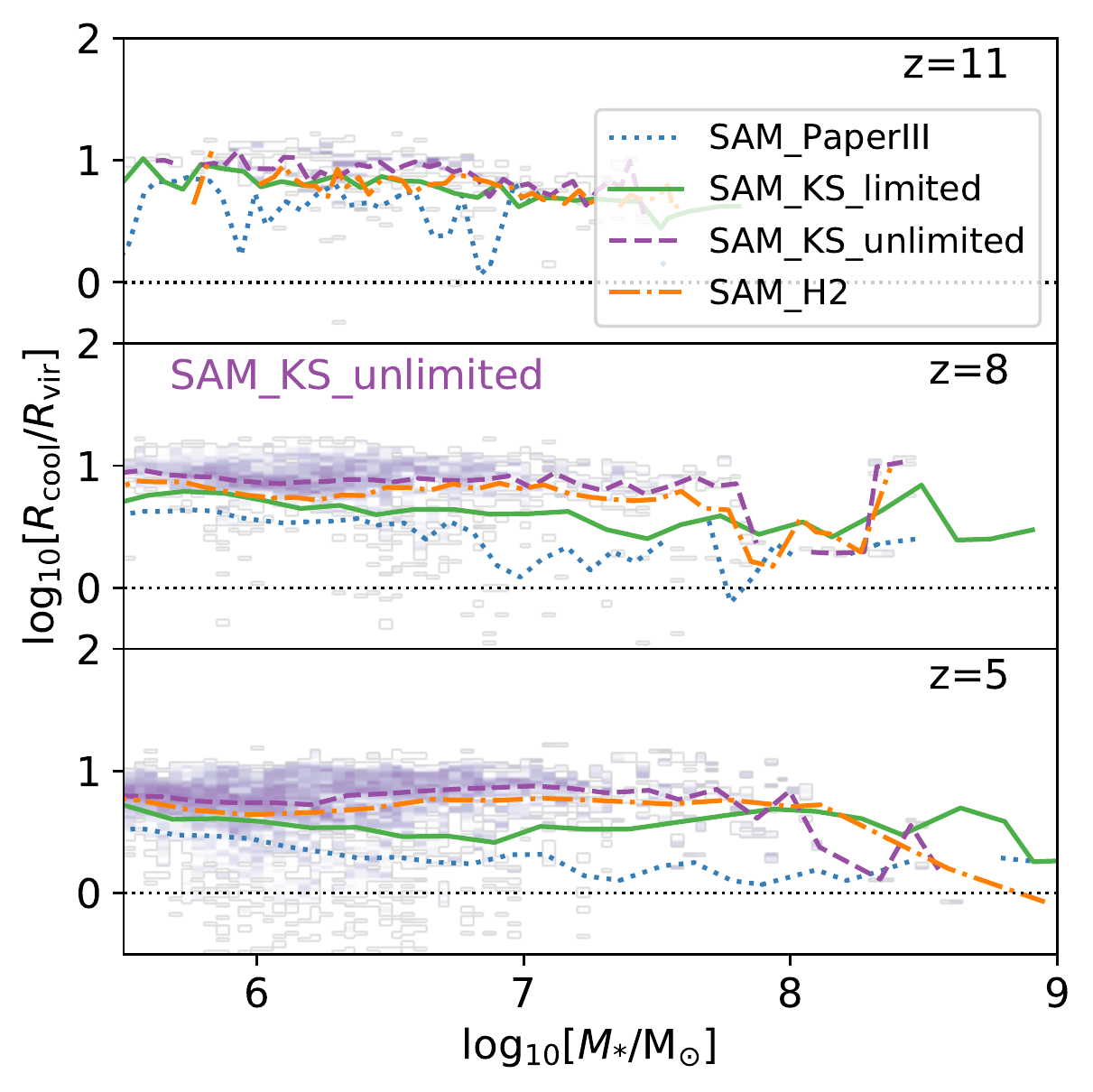}	
	\end{minipage}
	%	\vspace*{-3.8mm}
	\caption{\label{fig:rcool}The ratio of the cooling radius to the virial radius as a function of stellar mass ($M_*$) from the \textit{SAM\_PaperIII} (dotted line), \textit{SAM\_KS\_limited} (solid line), \textit{SAM\_KS\_unlimited} (dashed line) and \textit{SAM\_H2} (dash-dotted line) results at $z=11-5$. Lines represent the median while the 2D histogram shows the distribution in \textit{SAM\_KS\_unlimited}. Note that at each redshift, only galaxies with $M_\mathrm{vir}>10^9\mathrm{M}_\odot$ are considered and, in order to expand the sample size, we include objects from 7 consecutive snapshots (${\sim}80$Myr). Galaxies above the horizontal dotted line are identified as in the rapid cooling regime.}
\end{figure}

We note that when feedback is included, semi-analytic modelling of dwarf galaxies still suffers from these two factors. First, in order to demonstrate that most high-redshift dwarf galaxies in the SAM are still identified as in the rapid cooling regime when reionization and supernova feedback are included, we calculate the cooling radius, $R_\mathrm{cool}$, at which the time-scale of thermal cooling is equal to the halo dynamical time in the SAM and show the ratio of the cooling radius to the virial radius ($R_\mathrm{cool}/R_\mathrm{vir}$) calculated from the models discussed in this work in Fig. \ref{fig:rcool}. Note that gas within the cooling radius is considered to have reached hydrostatic equilibrium and cool thermally if $R_\mathrm{cool}<R_\mathrm{vir}$. However, in the case of a large cooling radius (i.e. $R_\mathrm{cool}>R_\mathrm{vir}$), the infalling gas will not be able to form stable shocks or remain in hydrostatic equilibrium. Accordingly, all of the accreted gas directly collapses into the central regions as free-fall. From Fig. \ref{fig:rcool}, we see that most low-mass galaxies discussed in this work are considered to be in the rapid cooling regime. 

\begin{figure*}
	\begin{minipage}{0.9\textwidth}
		\centering
		\includegraphics[width=\textwidth]{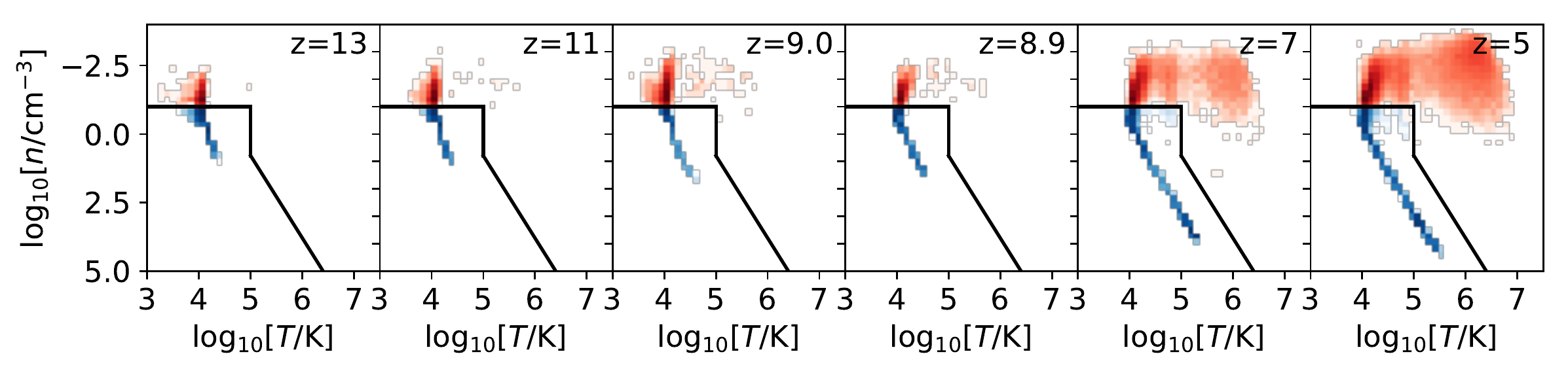}\\		\includegraphics[width=\textwidth]{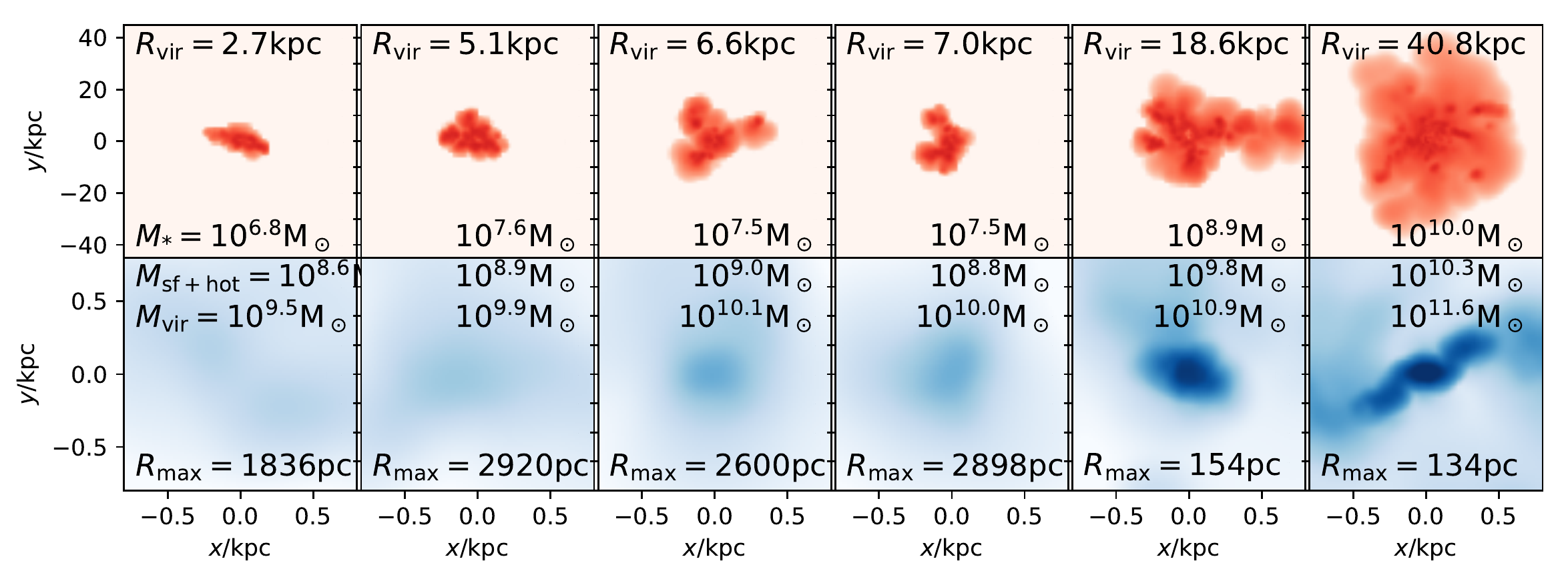}	
	\end{minipage}
	\caption{\label{fig:2dprofile} Profiles of the most massive $z=5$ halo in the \textit{WTHERM} {\smaug} simulation at $z=13-5$. \textit{Top panels:} gas density--temperature phase diagram. The black solid lines split the gas particles into star-forming (blue; inside the lower left region) and non-star-forming gas (red). \textit{Bottom panels:} face-on projections of the non-star-forming (top) and star-forming (bottom) gas particles. The stellar mass, gas mass, virial mass, virial radius and the radius of maximum rotation of this halo are shown in the bottom panels.}

	\begin{minipage}{\textwidth}
		\centering
		\includegraphics[width=0.495\textwidth]{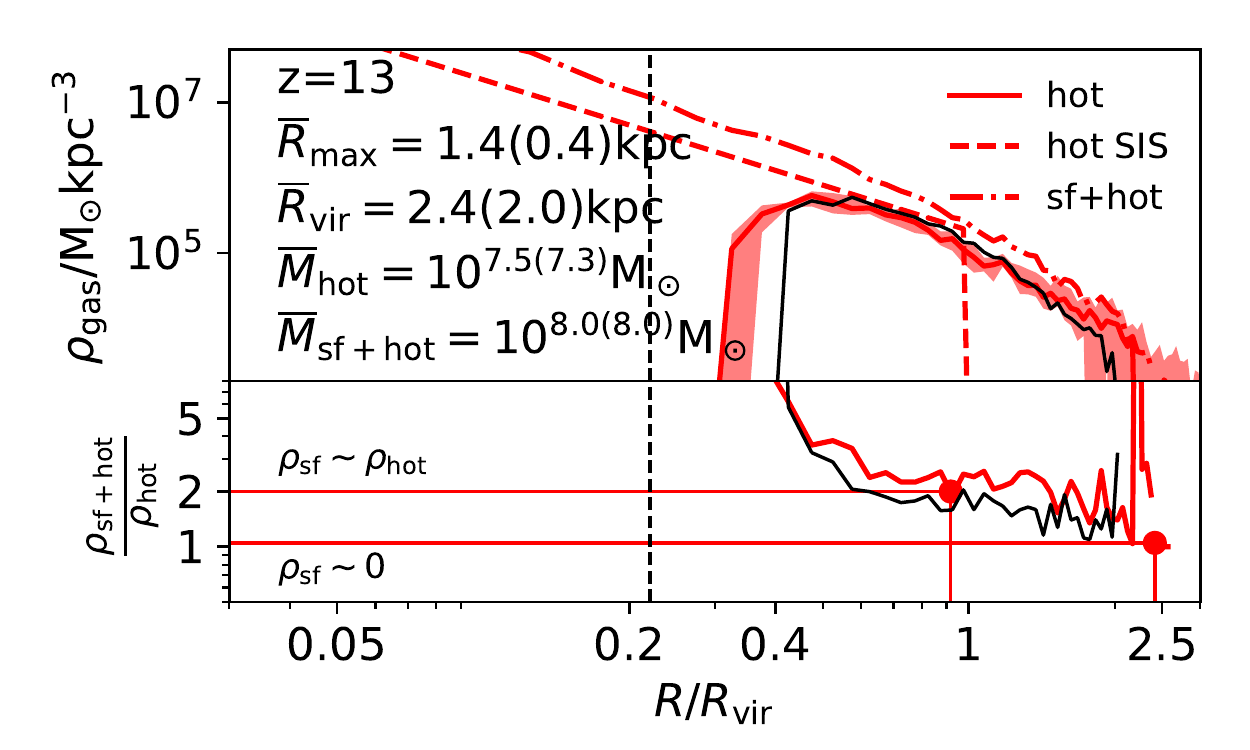}	\includegraphics[width=0.495\textwidth]{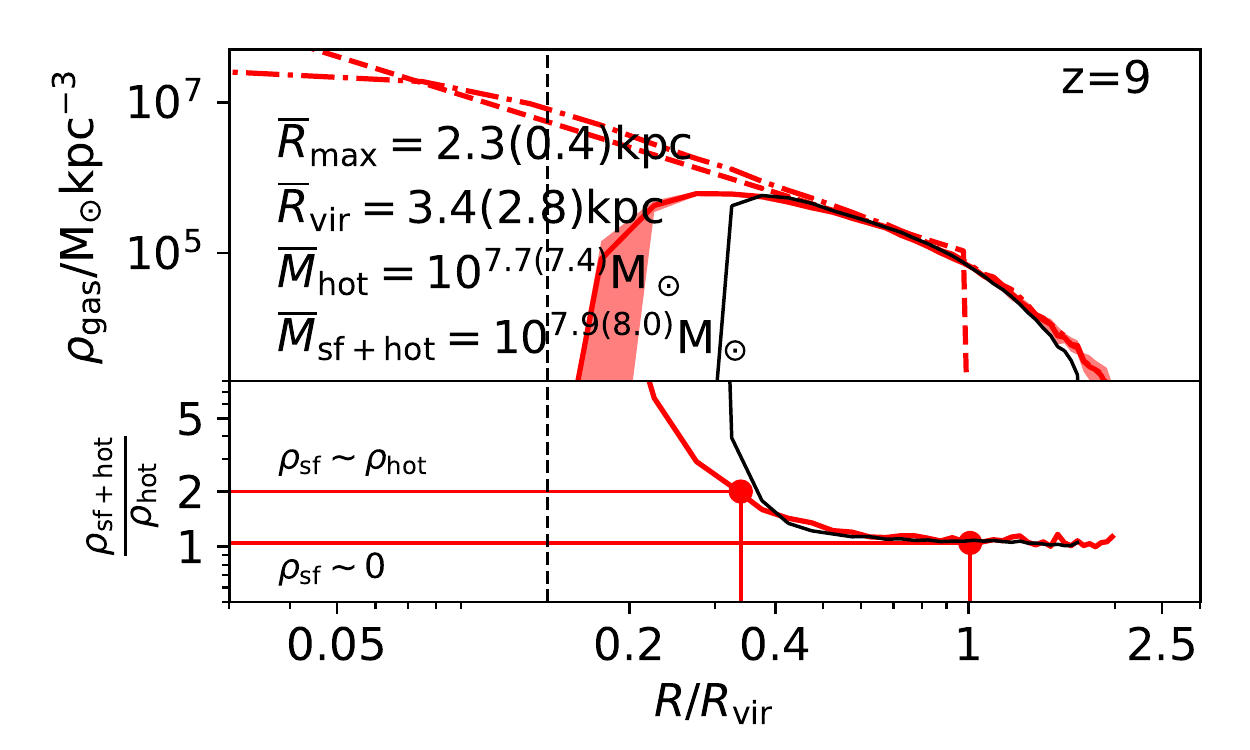}\\	\vspace*{-10mm}
		\includegraphics[width=0.495\textwidth]{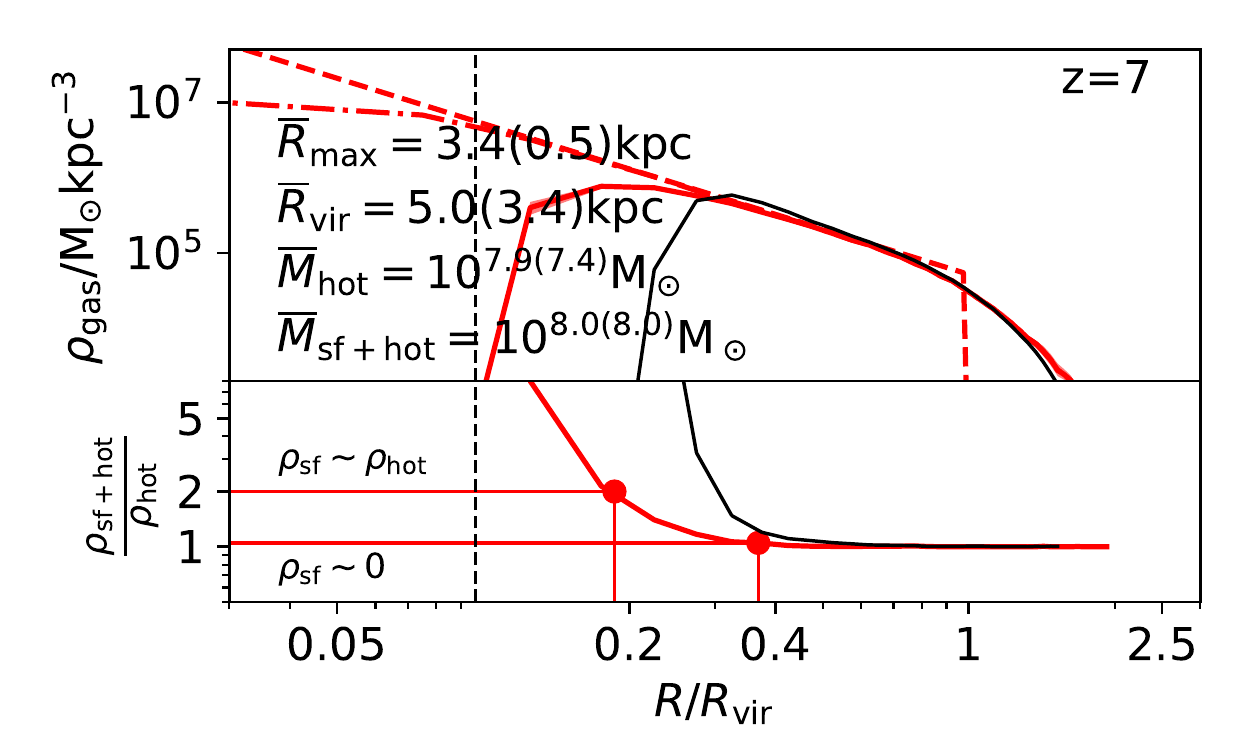}		\includegraphics[width=0.495\textwidth]{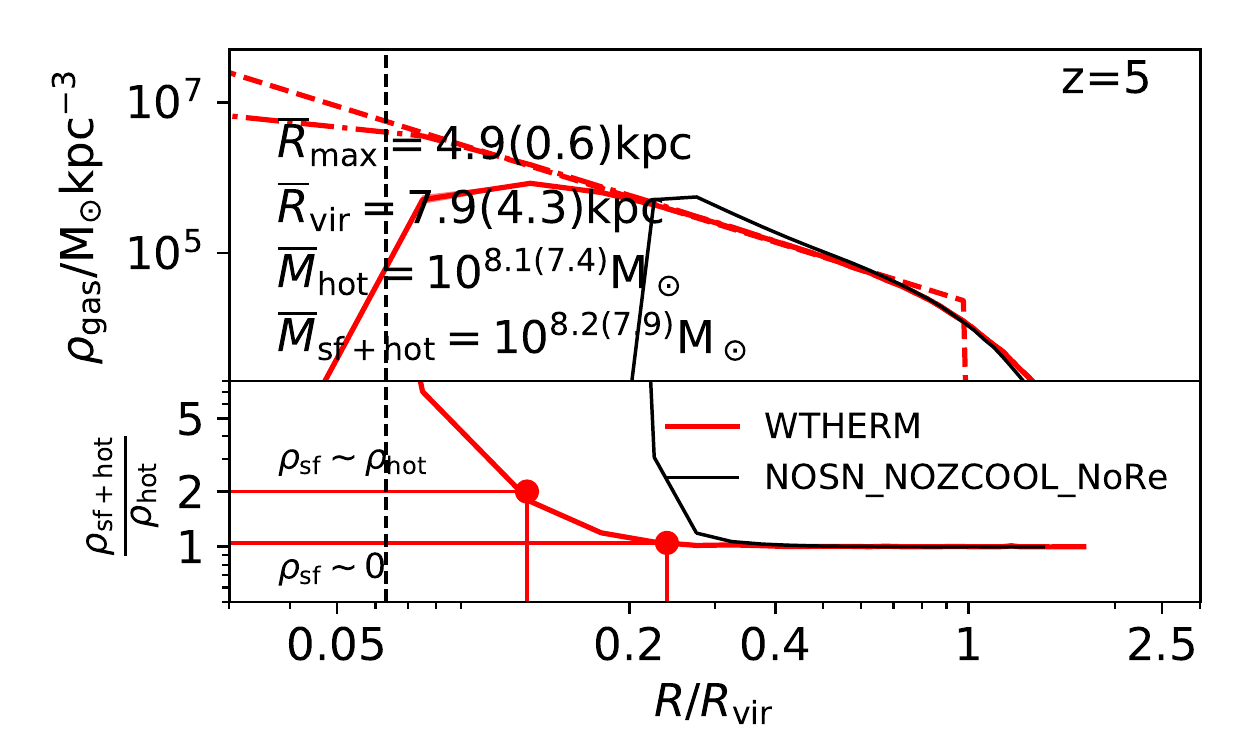}\vspace*{-3.5mm}
	\end{minipage}
	\caption{\label{fig:1dprofile}Gas profiles of galaxies with $M_*{\sim}10^{7\pm0.5}\rm{M}_\odot$ at $z=13-5$ in the \textit{WTHERM} {\smaug} simulation (red thick lines). In each panel, \textit{top:} the median radial density profiles of all gas (i.e. star-forming and non-star-forming gas; dash-dotted line) and the non-star-forming gas (solid line). Lines with shaded regions represent the median and 95 per cent confidence intervals around the median using 100000 bootstrap re-samples of the non-star-forming gas profile. The median SIS profile assumed in the SAM is calculated using the same amount of non-star-forming gas, and is indicated with the red dashed line; \textit{bottom:} the ratio of the density profiles of all gas to non-star-forming gas. The radius, within which the SIS gas is able to reach the centre through free-fall after one time step in the SAM, is indicated with thin vertical dashed lines. The radii, where the star-forming gas is as dense as the non-star-forming gas (i.e. $\rho_\mathrm{sf}\sim\rho_\mathrm{hot}$) or becomes deficient (i.e. $\rho_\mathrm{sf}\sim0$), are indicated with thin solid lines. The median virial radius, masses of non-star-forming and all gas mass are shown in the bottom right corner of each panel. The results of galaxies with $M_*{\sim}10^{7\pm0.5}\rm{M}_\odot$ in the \textit{NOSN\_NOZCOOL\_NoRe} are indicated with black thin lines for comparison, and their median properties are given in parenthesis.}
\end{figure*}

Next we show the evolution, in terms of the density--temperature phase and spatial distributions of star-forming and non-star-forming gas particles (identified using the algorithm described in \citetalias{qin2018}), of the most massive halo in the \textit{WTHERM} {\smaug} simulation identified at $z=5$ as an example in Fig. \ref{fig:2dprofile}, and discuss the gas density profile of galaxies with $M_*{\sim}10^{7\pm0.5}\rm{M}_\odot$ in Fig. \ref{fig:1dprofile}. We see that, compared to the \textit{NOSN\_NOZCOOL\_NoRe} simulation where heating from supernova (and reionization) is not included, galaxies within the same stellar mass range are hosted by larger haloes with more gas particles identified as non-star-forming when the feedback is considered. However, the total gas mass does not change significantly, indicating suppressions of baryonic mass and self-regulation of star formation. Moreover, although the star-forming regions become relatively smaller in \textit{WTHERM}, they still possess a large dispersion at high redshift. This can also be observed from the large radius of the maximum rotation ($R_\mathrm{max}$) of the most massive halo, which suggests the necessity of an enhanced inflow rate between the circum-galactic medium and ISM at earlier times.

More accurate semi-analytic modelling of gas accretion should not only distinguish the hot- and cold-mode inflows with gas reaching the star-forming disc on different time-scales \citep{Keres2005,Keres2009,Cattaneo2017}, but also account for the larger disc size at higher redshifts. We consider these as a future project with a more complete cooling function implemented. For the purpose of accurately capturing the gas transition time-scale using the current rapid cooling prescription, in \citetalias{qin2018}, we proposed to change the cooling efficiency when galaxies are identified in this regime. We introduced a maximum cooling factor, $\kappa_\mathrm{cool}$. This was used to modulate the gas transition time-scale based on the time-scale of free-fall, $t_\mathrm{transition}=\kappa_\mathrm{cool}^{-1}t_\mathrm{dyn}$, so that the overestimated collapse rate from the assumed SIS density profile and the underestimation due to the longer inflow path before the transition of gas reservoirs can be balanced. In this work, we adopt this modification by incorporating the following form of the transition time-scale.
\begin{equation}\label{eq:cooling}
t_\mathrm{transition}=\max\left[t_\mathrm{transition}^\mathrm{min}, \alpha_\mathrm{transition}\left(\dfrac{1+z}{6}\right)^{\beta_\mathrm{transition}}\right],
\end{equation}
where $t_\mathrm{transition}^\mathrm{min}$ is set to be $0.2t_\mathrm{dyn}$ following \citetalias{qin2018}. However, considering the transition radius between star-forming and non-star-forming gas changes due to feedback, $\alpha_\mathrm{transition}$ and $\beta_\mathrm{transition}$ are not expected to possess the same values as adopted in \citetalias{qin2018}\footnote{$\alpha_\mathrm{cool}\equiv\alpha_\mathrm{transition}^{-1}\times180\mathrm{Myr}=1$ and $\beta_\mathrm{cool}\equiv-1.5-\beta_\mathrm{transition}=1$ were utilized instead in \citetalias{qin2018}.}. Therefore, we leave them as free parameters and explore $t_\mathrm{transition}$ in this work.

\bsp
\label{lastpage}
\end{document}